# MATRIX RICCATI EQUATION SOLUTION OF THE 1D RADIATIVE TRANSFER EQUATION

**Barry D. Ganapol**
Department of Aerospace and Mechanical Engineering
University of Arizona
Tucson, AZ, 85721
Ganapol@cowboy.ame.arizona.edu

**Japan Patel**
Department of Mechanical and Aerospace Engineering,
The Ohio State University
Columbus, OH, 43210
patel.3545@osu.edu

## ABSTRACT

In recent years, the first author has developed three successful numerical methods to solve the 1D radiative transport equation yielding highly precise benchmarks. The second author has shown a keen interest in novel solution methodologies and an ability for their implementation. Here, we combine talents to generate yet another high precision solution, the Matrix Riccati Equation Method (MREM). MREM features the solution to two of the four matrix Riccati ODEs that arise from the interaction principle of particle transport. Through interaction coefficients, the interaction principle describes how particles reflect from- and transmit through- a single slab. On combination with Taylor series and doubling, a high quality numerical benchmark, to nearly seven places, is established.

## 1. INTRODUCTION

While the 1D radiative transfer equation is probably the most solved transport equation with numerous numerical and analytical solutions available, few solutions can claim to be an extreme benchmark. In context, an extreme benchmark means high confidence in six to seven places (seven to eight significant figures) of the numerical solution for angular intensity. In addition, 1D means exclusively a homogeneous plane parallel medium with a highly anisotropic scattering phase function represented either by a Legendre expansion or point-to-point discrete directions. The radiative transfer equation in question will simulate azimuthally integrated unpolarized radiation only. Thus, when referenced, the transport equation is for monochromatic- 1D- plane parallel- azithmuthally integrated- anistropically scattering- neutrons or unpolarized photons in a homogeneous medium. In addition, we limit our investigation to a perpendicular beam source impinging on the left (near) surface of a conservative ($\omega = 1$) slab medium in vacuum. Our intent is to develop a radiative transfer benchmark to extreme precision for anisotropically scattering slabs of small to modest optical thicknesses.

Our approach commonly falls under the category of the matrix-exponential method, which includes the discrete ordinates method, method of matrix operators and the matrix Riccati equations. These methods all share discretization of particle directions. The solutions include eigenvalue/eigenvector expansion, spatial finite difference and iterative sweeping in particle directions as well as doubling and adding via the interaction principle. For a through summary of the available literature, we refer the reader to the excellent work of Efremenko, et. al. [1].

### 1.1 The Matrix Riccati Equations (MREM) and the Interaction Principle

The interaction principle is the basic transport algorithm that defines Discrete Space Theory [2], which gives the solution to the transport equation with discretization only in direction $\mu$. The spatial variable $\tau$ remains continuous. The principle also goes under the name of invariant imbedding and is one of the first analytical methods developed to solve the radiative transfer equation [3]. The interaction principle is an alternative form of the discrete ordinate or SN equations avoiding the ad-hoc diamond difference approximation commonly found with spatial differencing. When combined with doubling, the interaction principle leads to a fast and precise numerical method. Recognizing that SN directional discretization is consistent, *i.e.*, as the number of discretizations become infinite, the exact solution results; offering an opportunity for convergence acceleration. Previously, convergence acceleration and doubling has been considered in a limited instance [4]. Here, convergence acceleration, through the Wynn-epsilon algorithm, enhances numerical efficiency of the interaction principle.

The matrix Riccati equations arise from the solution of the neutral particle transport equation in discrete ordinates form

$$\left[\mu_m \frac{\partial}{\partial \tau} + 1\right] I(\tau, \mu_m; N) = \sum_{m'=1}^{2N} \alpha_{m'} f(\mu_m, \mu_{m'}) I(\tau, \mu_{m'}; N), \ \tau_0 < \tau < \tau_1 \qquad (1a)$$

for directions $\mu_m$, $m = 1, 2, ..., 2N$. Thus, $\mu_m$ replaces $\mu$ in the angular intensity $I(\tau, \mu)$, where $\mu_m$ takes on both positive and negative values

$$\begin{cases} -\mu_m \\ \mu_{N+m} \equiv \mu_m \end{cases}, \ m = 1, ..., N, \ \ \mu_m > 0. \qquad (1b)$$

An $N^{\text{th}}$ order double Gauss quadrature approximates the collision integral over the half-range intervals [-1,0], [0,1] as

$$\int_{-1}^{1} d\mu' f(\mu,\mu') I(\tau,\mu') = \qquad , \qquad \text{(1c)}$$
$$= \sum_{m'=1}^{2N} \alpha_{m'} f(\mu_{m'},\mu) I(\tau,\mu_{m'};N) + E_G(\tau,\mu;N),$$

where $E_G(\tau,\mu;N)$ is the quadrature error [5]

$$E_G(\tau,\mu;N) \sim \left[\frac{e^2}{2^{10}\pi}\right]^N \frac{M(\tau,\mu)}{N^{4N}}. \qquad \text{(1d)}$$

For an integrand with at least 4*N* continuously bounded derivatives, the Gauss quadrature converges to its integral and therefore one can show the exact solution,

$$I(\tau,\mu) = I(\tau,\mu;N) + e(\tau,\mu;N),$$

with error $e(\tau,\mu;N)$, is the limit

$$I(\tau,\mu) = \lim_{N\to\infty} I(\tau,\mu;N). \qquad \text{(1e)}$$

The limit plays a prominent role in the application of convergence acceleration applied below, but before doing so, we must find the approximate intensity $I(\tau,\mu;N)$.

To form Eqs(1) as a vector equation, one defines the vectors and matrices (suppressing *N*)

$$\begin{aligned}
&\boldsymbol{I}^{\mp}(\tau) \equiv \begin{bmatrix} I_{\left\{\begin{smallmatrix}1\\N+1\end{smallmatrix}\right\}}(\tau) & I_{\left\{\begin{smallmatrix}2\\N+2\end{smallmatrix}\right\}}(\tau) & \ldots. & I_{\left\{\begin{smallmatrix}N\\2N\end{smallmatrix}\right\}}(\tau) \end{bmatrix}^T \\
&\boldsymbol{M} \equiv diag\{\mu_m\} \\
&\boldsymbol{C}^{\pm\pm}\boldsymbol{W} \equiv \begin{bmatrix} \boldsymbol{C}^{--} & \boldsymbol{C}^{-+} \\ \boldsymbol{C}^{+-} & \boldsymbol{C}^{++} \end{bmatrix} \boldsymbol{W} = \left\{\alpha_m f(\pm\mu_{m'},\pm\mu_m); j,m = 1,\ldots,N\right\} \\
&\boldsymbol{W} \equiv diag\{\alpha_m\},
\end{aligned} \qquad \text{(2a-d)}$$

where

$$I(\tau,\pm\mu_m;N) \equiv I_m^{\pm}(\tau) \qquad \text{(3a)}$$

to give

$$\frac{d\boldsymbol{I}^{\pm}(\tau)}{d\tau} = \mp \boldsymbol{M}^{-1}\left(\boldsymbol{I}_N - \boldsymbol{C}^{\pm\pm}\boldsymbol{W}\right)\boldsymbol{I}^{\pm}(\tau) \pm \boldsymbol{M}^{-1}\boldsymbol{C}^{\pm\mp}\boldsymbol{W}\boldsymbol{I}^{\mp}(\tau). \tag{3b}$$

$\boldsymbol{I}_N$ is the size $N$ matrix identity, not to be confused with intensity.

Therefore, by stacking the vectors in directions $\pm\mu_m$

$$\boldsymbol{I}(\tau) \equiv \begin{bmatrix} \boldsymbol{I}^{-}(\tau) \\ \boldsymbol{I}^{+}(\tau) \end{bmatrix},$$

we have

$$\frac{d\boldsymbol{I}(\tau)}{d\tau} = \boldsymbol{A}\boldsymbol{I}(\tau), \tag{4a}$$

with

$$\boldsymbol{A} \equiv \begin{bmatrix} \boldsymbol{\alpha} & -\boldsymbol{\beta} \\ \boldsymbol{\beta} & -\boldsymbol{\alpha} \end{bmatrix}. \tag{4b}$$

Since from scattering invariance,

$$\boldsymbol{C}^{+-} = \boldsymbol{C}^{-+},$$

there results

$$\begin{aligned} \boldsymbol{\alpha} &\equiv \boldsymbol{M}^{-1}\left(\boldsymbol{I}_N - \boldsymbol{C}^{++}\boldsymbol{W}\right) \\ \boldsymbol{\beta} &\equiv \boldsymbol{M}^{-1}\boldsymbol{C}^{+-}\boldsymbol{W}. \end{aligned} \tag{4c,d}$$

Equations (1) and (4a) are subject to the incoming intensity at $\tau = \tau_0$ and no source incoming intensity at $\tau = \tau_1$

$$\boldsymbol{I}^{+}(\tau_0) \equiv \boldsymbol{g},\ \boldsymbol{I}^{-}(\tau_1) \equiv \boldsymbol{0} \tag{1f}$$

to complete the problem statement. $\boldsymbol{g}$ is the angular intensity for a perpendicular incoming beam on the near surface $\tau_0$ specified later. Since we consider a conservative medium, there is no re- radiation making the medium source less.

At this point in our development, there are several paths forward to achieve a benchmarking goal. For your interest, APPENDIX A includes brief descriptions of three successful approaches. All methods referenced generate benchmarks to seven places to both the HAZE L 82 and Cloud C1 300 term scattering kernels [6]. We now give the details of the Matrix Riccati Equation Method (MREM) of solution featuring doubling, and offering an entirely different formalism than previous solutions.

### 1.1.1 Interaction principle

Consider a slab layer of thickness $\tau - \tau_0$ as shown in Fig. 1. Rather than seek an analytical exponential eigenvalue/eigenfunction based solution to Eq(4a), equally appropriate since $\boldsymbol{A}$ is independent of $\tau$, we derive the solution $\boldsymbol{P}(\tau,\tau_0)$ to the corresponding fundamental matrix equation

$$\frac{d\boldsymbol{P}(\tau,\tau_0)}{d\tau} = \boldsymbol{A}\boldsymbol{P}(\tau,\tau_0);\ \boldsymbol{P}(\tau_0,\tau_0) = \boldsymbol{I}_N \tag{5a}$$

by letting

$$\boldsymbol{I}(\tau) = \boldsymbol{P}(\tau;\tau_0)\boldsymbol{I}(\tau_0). \tag{5b}$$

$\boldsymbol{P}(\tau,\tau_0)$ is also called the transition matrix and remains valid even when the $A$ matrix

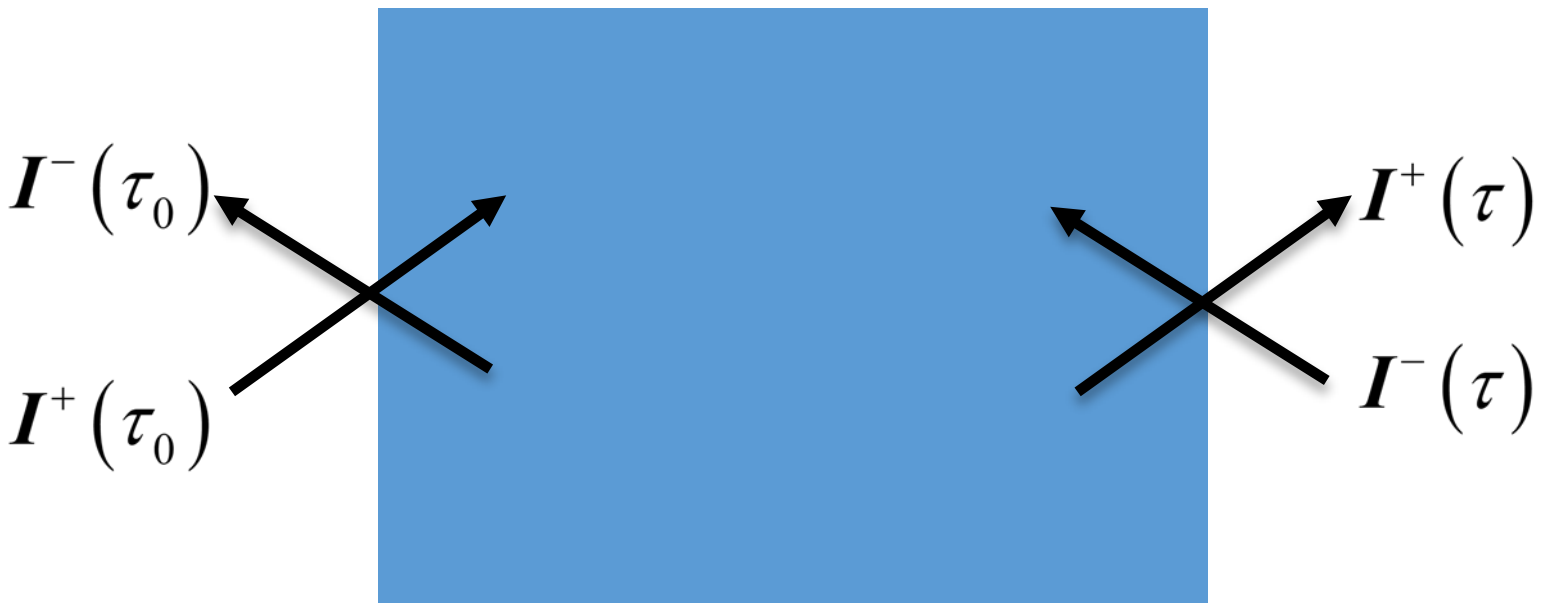

**Fig. 1. Single Slab.**

depends upon $\tau$. Therefore, Eqs(5) are the starting point for consideration of variable scattering across the slab (a future effort).

Recalling that the superscript of the intensity indicates direction of particle motion, we write the stacked vectors at the boundaries shown in Fig. 1 as

$$\boldsymbol{I}(\tau_0) = \begin{bmatrix} \boldsymbol{I}^-(\tau_0) \\ \boldsymbol{I}^+(\tau_0) \end{bmatrix}$$
$$\boldsymbol{I}(\tau) = \begin{bmatrix} \boldsymbol{I}^-(\tau) \\ \boldsymbol{I}^+(\tau) \end{bmatrix}. \tag{6a,b}$$

The interaction principle says a homogeneous slab will reflect and transmit particles entering either face equally. Thus, if $\boldsymbol{R}(\tau-\tau_0)$ and $\boldsymbol{T}(\tau-\tau_0)$ are the reflectance and transmittance matrix coefficients of the homogeneous medium depending on the medium thickness and scattering and absorbing properties, we can the write from physical reasoning

$$\begin{aligned} \boldsymbol{I}^-(\tau_0) &= \boldsymbol{R}(\tau-\tau_0)\boldsymbol{I}^+(\tau_0) + \boldsymbol{T}(\tau-\tau_0)\boldsymbol{I}^-(\tau) \\ \boldsymbol{I}^+(\tau) &= \boldsymbol{T}(\tau-\tau_0)\boldsymbol{I}^+(\tau_0) + \boldsymbol{R}(\tau-\tau_0)\boldsymbol{I}^-(\tau), \end{aligned} \tag{7a}$$

or in vector form

$$\begin{bmatrix} \boldsymbol{I}^-(\tau_0) \\ \boldsymbol{I}^+(\tau) \end{bmatrix} = \begin{bmatrix} \boldsymbol{T}(\tau-\tau_0) & \boldsymbol{R}(\tau-\tau_0) \\ \boldsymbol{R}(\tau-\tau_0) & \boldsymbol{T}(\tau-\tau_0) \end{bmatrix} \begin{bmatrix} \boldsymbol{I}^-(\tau) \\ \boldsymbol{I}^+(\tau_0) \end{bmatrix}, \tag{7b}$$

which is the interaction principle. Thus, the exiting intensity at surface $\tau_0$ results from reflection from the incoming at $\tau_0$ and that transmitted from the far surface at $\tau$. Likewise, the intensity exiting the surface at $\tau$ results from that reflected from the incoming at $\tau$ and that transmitted from the near surface at $\tau_0$.

The interaction principle separates the slab influence on the particles within the slab into reflection and transmission. The RHS is the source of particles into the slab and the LHS is the consequence of particle interaction within the slab, or output response. Hence, we call the matrix connecting the output to the input the response matrix

$$\boldsymbol{Q}(\tau-\tau_0) = \begin{bmatrix} \boldsymbol{T}(\tau-\tau_0) & \boldsymbol{R}(\tau-\tau_0) \\ \boldsymbol{R}(\tau-\tau_0) & \boldsymbol{T}(\tau-\tau_0) \end{bmatrix}. \tag{8a}$$

to give the interaction principle for a homogeneous slab as

$$\begin{bmatrix} \boldsymbol{I}^-(\tau_0) \\ \boldsymbol{I}^+(\tau) \end{bmatrix} = \boldsymbol{Q}(\tau-\tau_0) \begin{bmatrix} \boldsymbol{I}^-(\tau) \\ \boldsymbol{I}^+(\tau_0) \end{bmatrix}. \tag{8b}$$

By construction

$$\begin{aligned} \boldsymbol{R}(0) &= \boldsymbol{0} \\ \boldsymbol{T}(0) &= \boldsymbol{I}_N . \end{aligned} \tag{8c}$$

The response matrix is block symmetric and depends upon slab thickness and material scattering and absorption properties.

Henceforth, to simplify without loss of generality, we let $\tau_0 = 0$; and therefore in the context used here, $\tau$ is the slab thickness. In addition, when obvious, we suppress the interaction coefficient dependence on slab thickness.

The interaction principle is an exact formulation of particle transport, as much as the radiative transfer equation is, once we know the reflectance and transmittance, which now becomes our task.

### 1.1.2 The matrix Riccati equations

Re-arranging the vector equation Eq(8b) gives

$$\begin{bmatrix} \boldsymbol{T} & \boldsymbol{0} \\ -\boldsymbol{R} & \boldsymbol{I}_N \end{bmatrix} \boldsymbol{I}(\tau) = \begin{bmatrix} -\boldsymbol{I}_N & \boldsymbol{R} \\ \boldsymbol{0} & \boldsymbol{T} \end{bmatrix} \boldsymbol{I}(0) \cdot \tag{9a}$$

or

$$\boldsymbol{I}(\tau) = \begin{bmatrix} \boldsymbol{T} & \boldsymbol{0} \\ -\boldsymbol{R} & \boldsymbol{I}_N \end{bmatrix}^{-1} \begin{bmatrix} -\boldsymbol{I}_N & \boldsymbol{R} \\ \boldsymbol{0} & \boldsymbol{T} \end{bmatrix} \boldsymbol{I}(0) \cdot \tag{9b}$$

From Schur's complement

$$\begin{bmatrix} \boldsymbol{T} & \boldsymbol{0} \\ -\boldsymbol{R} & \boldsymbol{I}_N \end{bmatrix}^{-1} = \begin{bmatrix} \boldsymbol{T}^{-1} & \boldsymbol{0} \\ \boldsymbol{R}\boldsymbol{T}^{-1} & \boldsymbol{I}_N \end{bmatrix},$$

and therefore

$$\boldsymbol{I}(\tau) = \begin{bmatrix} \boldsymbol{T}^{-1} & -\boldsymbol{T}^{-1}\boldsymbol{R} \\ \boldsymbol{R}\boldsymbol{T}^{-1} & \boldsymbol{T} - \boldsymbol{R}\boldsymbol{T}^{-1}\boldsymbol{R} \end{bmatrix} \boldsymbol{I}(0) \cdot \tag{10}$$

Then, from Eq(5b)

$$\boldsymbol{P}(\tau) = \begin{bmatrix} \boldsymbol{T}^{-1} & -\boldsymbol{T}^{-1}\boldsymbol{R} \\ \boldsymbol{R}\boldsymbol{T}^{-1} & \boldsymbol{T} - \boldsymbol{R}\boldsymbol{T}^{-1}\boldsymbol{R} \end{bmatrix} \cdot \tag{11}$$

Finally, the following four Matrix Riccati differential equations result for ***T*** and ***R*** directly from Eq(5a):

$$\begin{bmatrix} \frac{d\boldsymbol{T}^{-1}}{d\tau} & \frac{d\boldsymbol{T}^{-1}}{d\tau}\boldsymbol{R}-\boldsymbol{T}^{-1}\frac{d\boldsymbol{R}^{-1}}{d\tau} \\ \frac{d\boldsymbol{R}^{-1}}{d\tau}\boldsymbol{T}^{-1}+\boldsymbol{R}\frac{d\boldsymbol{T}^{-1}}{d\tau}+ & \frac{d\boldsymbol{T}^{-1}}{d\tau}-\frac{d}{d\tau}\left(\boldsymbol{R}\boldsymbol{T}^{-1}\boldsymbol{R}\right) \end{bmatrix}=\begin{bmatrix} \boldsymbol{\alpha} & -\boldsymbol{\beta} \\ \boldsymbol{\beta} & -\boldsymbol{\alpha} \end{bmatrix}\begin{bmatrix} \boldsymbol{T}^{-1} & -\boldsymbol{T}^{-1}\boldsymbol{R} \\ \boldsymbol{R}\boldsymbol{T}^{-1} & \boldsymbol{T}-\boldsymbol{R}\boldsymbol{T}^{-1}\boldsymbol{R} \end{bmatrix},$$

which give by equating corresponding block matrix components (after some algebra)

$$\begin{aligned} \frac{d\boldsymbol{R}}{d\tau} &= \boldsymbol{T}\boldsymbol{\beta}\boldsymbol{T} \\ \frac{d\boldsymbol{T}}{d\tau} &= \boldsymbol{R}\boldsymbol{\beta}\boldsymbol{T}-\boldsymbol{\alpha}\boldsymbol{T} \\ \frac{d\boldsymbol{R}}{d\tau} &= \boldsymbol{\beta}-\boldsymbol{R}\boldsymbol{\alpha}-\boldsymbol{\alpha}\boldsymbol{R}+\boldsymbol{R}\boldsymbol{\beta}\boldsymbol{R} \\ \frac{d\boldsymbol{T}}{d\tau} &= \boldsymbol{T}\boldsymbol{\beta}\boldsymbol{R}-\boldsymbol{T}\boldsymbol{\alpha}. \end{aligned} \qquad (12a\text{-}d)$$

In addition, there are two obvious identities

$$\begin{aligned} \boldsymbol{T}\boldsymbol{\beta}\boldsymbol{T} &= \boldsymbol{\beta}-\boldsymbol{R}\boldsymbol{\alpha}-\boldsymbol{\alpha}\boldsymbol{R}+\boldsymbol{R}\boldsymbol{\beta}\boldsymbol{R} \\ \boldsymbol{R}\boldsymbol{\beta}\boldsymbol{T}-\boldsymbol{\alpha}\boldsymbol{T} &= \boldsymbol{T}\boldsymbol{\beta}\boldsymbol{R}-\boldsymbol{T}\boldsymbol{\alpha}. \end{aligned} \qquad (12e,f)$$

Next, we need to determine analytical expressions for interaction coefficients ***R*** and ***T***.

## 2. TAYLOR SERIES REPRESENTATIONS FOR *R* AND *T*

At this point, there are numerous ways to numerically evaluate the interaction coefficients. Possibilities include the multiple collision technique [7], where one partitions intensity into particle collisions whose sum is the intensity at $\tau$ and $\mu$. This approach becomes analytically unwieldly with increasing number of collisions [1]. Another method forms integral equations for Eqs(12) and proceeds iteratively [8], which is essentially the Taylor series (TS) formulation. Also, integration by Runge-Kutta methods have been applied to this non-linear set of ODEs [9].

Here, we will follow the TS representation in two distinct forms, direct and indirect. The direct form comes from Eqs(12c,d), where the interaction coefficients have a TS representation in $\tau$ even though equations are non-linear. The indirect form comes from a product transform rendering Eqs(12) linear in each factor of the product. A TS then applies to each factor with the resulting product also a TS.

## 2.1 Direct Taylor Series Representation

Even though the Matrix Riccati equations are quadratically nonlinear, analytical solutions, possibly not in closed form, are possible through Taylor series.

### 2.1.1 Reflectance

We now focus on solving the matrix Riccati equation Eq(12c) for the reflectance $\boldsymbol{R}$ since it is uncoupled. The most convenient solution is the TS formulation

$$\boldsymbol{R}(\tau)=\sum_{n=0}^{\infty}\boldsymbol{R}_n\tau^n, \tag{13a}$$

where

$$\boldsymbol{R}_n(\tau)\equiv\frac{1}{n!}\boldsymbol{R}^{(n)}(\tau)\Big|_{\tau=0}.$$

Then, applying $n$-1 derivatives to Eq(12c) and dividing by $n$!, results in

$$n\boldsymbol{R}_n(\tau)=\boldsymbol{\beta}\delta_{n,1}-\boldsymbol{R}_{n-1}(\tau)\boldsymbol{\alpha}-\boldsymbol{\alpha}\boldsymbol{R}_{n-1}(\tau)+\frac{1}{n!}\frac{d^{n-1}}{d\tau^{n-1}}\boldsymbol{R}(\tau)\boldsymbol{\beta}\boldsymbol{R}(\tau). \tag{13b}$$

The last term comes from Leibnitz's differentiation rule for scalars and vectors alike

$$\frac{1}{n!}\frac{d^{n-1}}{d\tau^{n-1}}f(\tau)g(\tau)=\sum_{l=1}^{n-1}f_{n-1-l}(\tau)g_l(\tau) \tag{14}$$

giving the recurrence for $\tau=0$

$$n\boldsymbol{R}_n=\boldsymbol{\beta}\delta_{n-1,0}-\boldsymbol{R}_{n-1}\boldsymbol{\alpha}-\boldsymbol{\alpha}\boldsymbol{R}_{n-1}+\sum_{l=1}^{n-1}\boldsymbol{R}_{n-1-l}\boldsymbol{\beta}\boldsymbol{R}_l. \tag{15}$$

### 2.1.2 Transmittance

Similarly, if

$$\boldsymbol{T}(\tau)=\sum_{n=0}^{\infty}\boldsymbol{T}_n\tau^n, \tag{16a}$$

by applying $n$-1 derivatives to Eq(12d), dividing by $n!$ and by application of Leibnitz's differentiation,

$$\frac{\boldsymbol{T}^{(n)}(\tau)}{n!} = \frac{1}{n!}\frac{d^{n-1}}{d\tau^{n-1}}\boldsymbol{T}(\tau)\boldsymbol{\beta}\boldsymbol{R}(\tau) - \frac{\boldsymbol{T}^{(n-1)}(\tau)}{n(n-1)!}\boldsymbol{\alpha},$$

or

$$n\boldsymbol{T}_n = \sum_{l=1}^{n-1}\boldsymbol{T}_{n-1-l}\boldsymbol{\beta}\boldsymbol{R}_l - \boldsymbol{T}_{n-1}\boldsymbol{\alpha}, \tag{16b}$$

where

$$\boldsymbol{T}_n \equiv \frac{1}{n!}\boldsymbol{T}^{(n)}(\tau)\Big|_{\tau=0}.$$

While the direct TS is certainly numerically adequate, it is of analytical interest to derive an alternative.

### 2.2 Indirect Taylor Series Representation

One can linearize Eqs(12) through the following transformations:

$$\boldsymbol{R}(\tau) \equiv \boldsymbol{Q}(\tau)\boldsymbol{P}^{-1}(\tau) \tag{B1}$$

$$\boldsymbol{T} = \boldsymbol{W}\boldsymbol{P}^{-1} \tag{B14}$$

(Do not to confuse $\boldsymbol{Q}$ with the response matrix). As one shows in APPENDIX B, $\boldsymbol{W}$ is unity and the corresponding TS coefficients are

$$\boldsymbol{R}_n = \boldsymbol{Q}_n - \sum_{l=1}^{n}\boldsymbol{P}_{n-l}\boldsymbol{R}_l. \tag{B12}$$

$$\boldsymbol{T}_n = \boldsymbol{I}_N\delta_{n0} - \sum_{l=0}^{n-1}\boldsymbol{P}_{n-l}\boldsymbol{T}_l. \tag{B19}$$

Numerically, the direct and indirect representations give identical results, but linearization is important for the theoretical connection between formulations.

## 3. NUMERICAL EVALUATION

We are now in position to establish numerical implementation. Built around Eq(8b) with $\tau_0 = 0$ and thickness $\tau_L$ expressed as

$$\begin{bmatrix} \boldsymbol{I}^-(0) \\ \boldsymbol{I}^+(\tau_L) \end{bmatrix} = \boldsymbol{Q}(\tau_L) \begin{bmatrix} \boldsymbol{I}^-(\tau_L) \\ \boldsymbol{I}^+(0) \end{bmatrix}, \tag{17}$$

our strategy is to first determine the interaction coefficients from their TS representations of Eqs(13a) and (16a). We follow this by application of the method of doubling to find the overall slab response matrix to give the outgoing intensity from the incoming intensity specified at slab boundaries according to Eq(17). In this way, one finds the exiting angular intensity distributions. The interior angular intensity distribution at $\tau$ comes from the method of adding responses of two complimentary slabs of thickness $\tau$ and $\tau_L$-$\tau$. In doing so, we find the intensity distribution at any point within the slab without requiring the unknown exiting intensity distributions as a pre-calculation. We believe this procedure to be knew. As will be discussed, each numerical procedure must include additional features to make MREM worthy of a benchmark.

### 3.1 Exiting Intensity by Doubling

The method of doubling begins with the interaction principle for a slab of thickness $\tau_{l\text{-}1}$ in block matrix form

$$\begin{bmatrix} \boldsymbol{I}_0^- \\ \boldsymbol{I}_{l-1}^+ \end{bmatrix} = \boldsymbol{Q}_{l-1} \begin{bmatrix} \boldsymbol{I}_{l-1}^- \\ \boldsymbol{I}_0^+ \end{bmatrix} = \begin{bmatrix} \boldsymbol{Q}_{l-1,1} & \boldsymbol{Q}_{l-1,2} \\ \boldsymbol{Q}_{l-1,2} & \boldsymbol{Q}_{l-1,1} \end{bmatrix} \begin{bmatrix} \boldsymbol{I}_{l-1}^- \\ \boldsymbol{I}_0^+ \end{bmatrix}. \tag{18}$$

(Note: We suppress the argument in the response matrix.)

Figure 2 shows the slab, called slab $l$-1 with intensities at boundaries at 0 and $\tau_{l\text{-}1}$. Equation (18) takes into account block symmetry of the reflectance and transmittance coefficients

$$\boldsymbol{Q}_{l-1,3} = \boldsymbol{Q}_{l-1,2} = \boldsymbol{R}_{l-1}, \;\; \boldsymbol{Q}_{l-1,4} = \boldsymbol{Q}_{l-1,1} = \boldsymbol{T}_{l-1}.$$

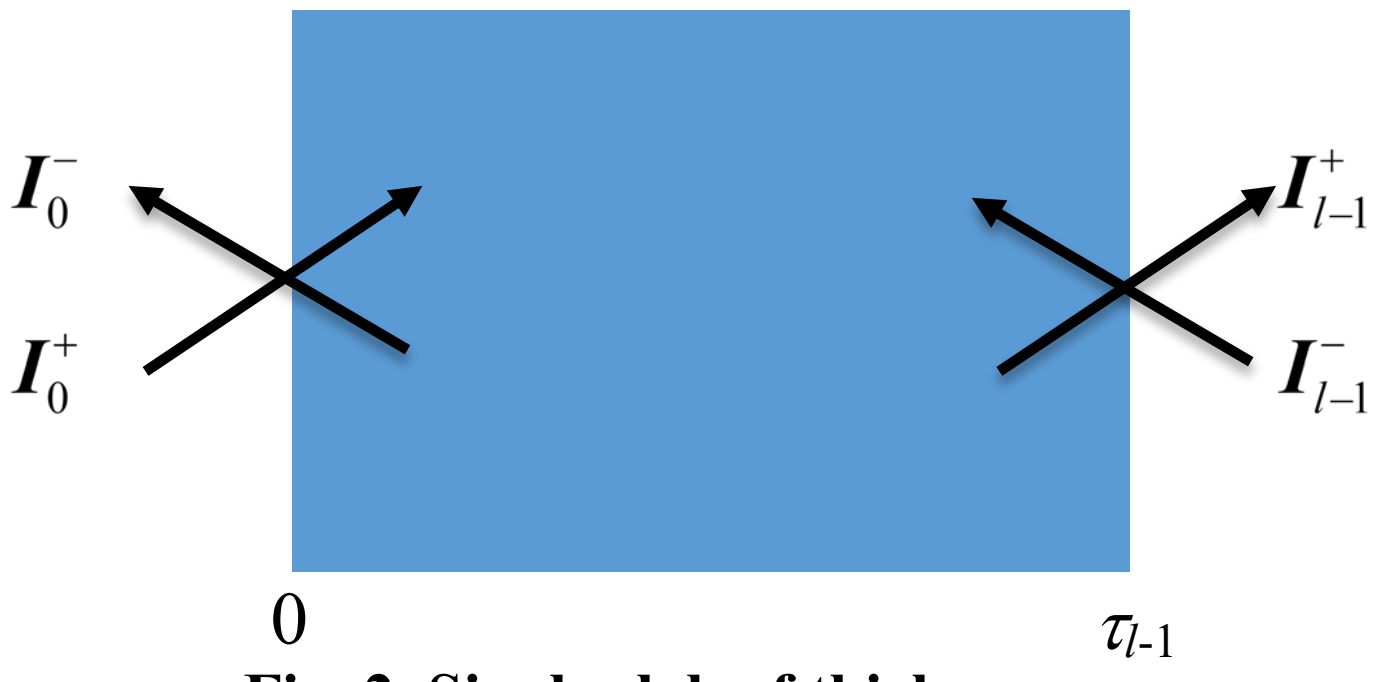

**Fig. 2. Single slab of thickness $\tau_{l\text{-}1}$.**

Note that a single slab can be a combination of many slabs whose equivalent response matrix comes from the following procedure.

Consider a slab of thickness $\tau_l$ - $\tau_{l\text{-}1}$ with different properties and interaction principle

$$\begin{bmatrix} I_{l-1}^- \\ I_l^+ \end{bmatrix} = \mathscr{R}_l \begin{bmatrix} I_l^- \\ I_{l-1}^+ \end{bmatrix} = \begin{bmatrix} T_l & R_l \\ R_l & T_l \end{bmatrix} \begin{bmatrix} I_l^- \\ I_{l-1}^+ \end{bmatrix}$$

joined to the first slab as shown in Fig. 3. Writing out the interaction principle

$$\begin{aligned} I_0^- &= Q_{l-1,1} I_{l-1}^- + Q_{l-1,2} I_0^+, \quad & I_l^+ &= R_l I_l^- + T_l I_{l-1}^+ \\ I_{l-1}^+ &= Q_{l-1,2} I_{l-1}^- + Q_{l-1,1} I_0^+, \quad & I_{l-1}^- &= T_l I_l^- + R_l I_{l-1}^+; \end{aligned} \tag{19}$$

and eliminating the intensities at the interface gives the combined response $Q_l$ for the composite slab $[0, \tau_l]$ and the exiting intensities

$$\begin{bmatrix} I_0^- \\ I_l^+ \end{bmatrix} = Q_l(\tau_l) \begin{bmatrix} I_l^- \\ I_0^+ \end{bmatrix}, \tag{20a}$$

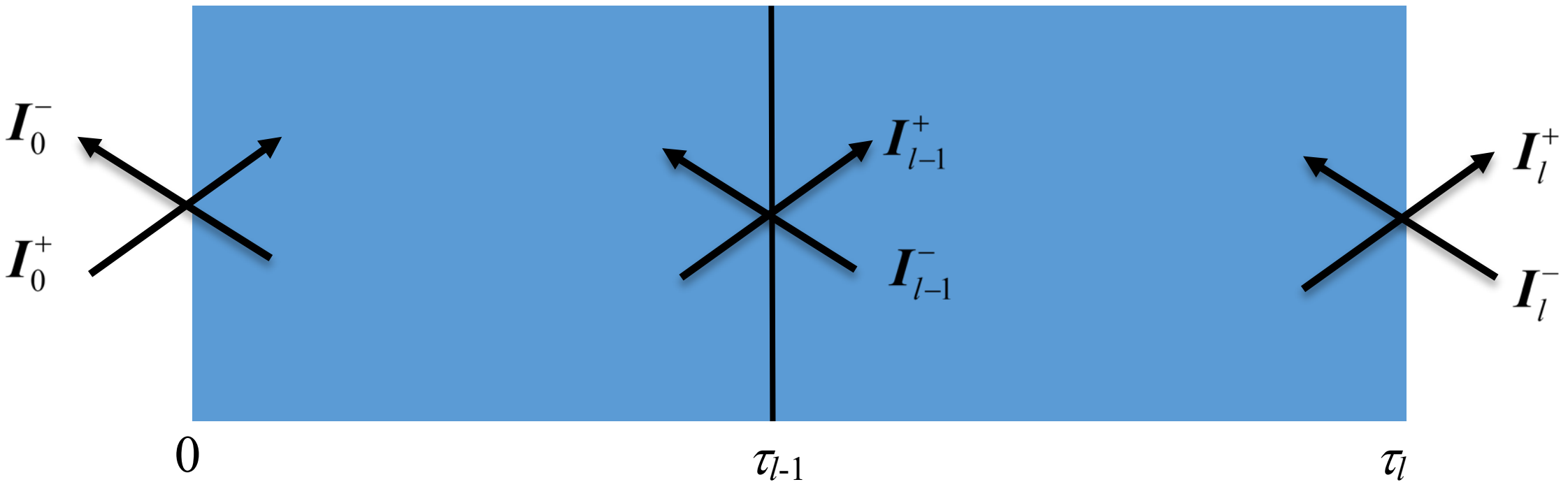

**Fig. 3. Two Slabs.**

from block recurrence

$$Q_l(\tau_l) \equiv \begin{bmatrix} Q_{l,1} & Q_{l,2} \\ Q_{l,3} & Q_{l,4} \end{bmatrix} = \begin{bmatrix} \mathbf{0} & Q_{l-1,1} \\ T_l & \mathbf{0} \end{bmatrix} U_l + \begin{bmatrix} \mathbf{0} & Q_{l-1,2} \\ R_l & \mathbf{0} \end{bmatrix}, \tag{20b}$$

with

$$\begin{aligned} w_{l-} &\equiv \begin{bmatrix} I_N & -Q_{l-1,2} \\ -R_l & I_N \end{bmatrix} \\ w_{l+} &\equiv \begin{bmatrix} \mathbf{0} & Q_{l-1,2} \\ T_l & \mathbf{0} \end{bmatrix} \\ U_l(\tau_l, \tau_l - \tau_{l-1}) &= w_{l-}^{-1} w_{l+}. \end{aligned} \tag{20c,d,e}$$

Finally, from the interaction principle in terms of slab *l* interaction coefficients,

$$\boldsymbol{Q}_l\left(\tau_{l-1}\right) \equiv \begin{bmatrix} \boldsymbol{T}_l & \boldsymbol{R}_l \\ \boldsymbol{R}_l & \boldsymbol{T}_l \end{bmatrix}.$$

Thus, the response for joined slabs of any thicknesses and properties results from recurrence for each of the four square matrix partitions of the response $\boldsymbol{Q}_l$. While not initially apparent, the composite response is also block symmetric. In addition, in deriving Eq(20b), an intermediate step gives the intensity at the interface [10]

$$\begin{bmatrix} \boldsymbol{I}_{l-1}^{+} \\ \boldsymbol{I}_{l-1}^{-} \end{bmatrix} = \boldsymbol{U}_l\left(\tau_l, \tau_l - \tau_{l-1}\right) \begin{bmatrix} \boldsymbol{I}_l^{-} \\ \boldsymbol{I}_0^{+} \end{bmatrix} \tag{21}$$

to be used below. We are ready to double.

Doubling arises from the definition of $\boldsymbol{P}_1(h)$ for a single slab (designated 1) from Eq(5b) with $\tau_0 = 0$ of thickness $h$

$$\boldsymbol{I}(h) = \boldsymbol{P}_1(h)\boldsymbol{I}(0) \tag{22}$$

relating incoming to outgoing intensities. $\boldsymbol{P}_1(h)$ acts as a propagator, propagating particles back and forth through the slab. For a composite of two identical slabs, each with interaction coefficients $\boldsymbol{R}(h)$and $\boldsymbol{T}(h)$, propagation represents a joint probably of particle transfer between sequential surfaces of the composite slabs, with the exiting intensities

$$\boldsymbol{I}(2h) = \boldsymbol{P}_1(h)\boldsymbol{I}(h) \tag{23a}$$

or from Eq(22)

$$\boldsymbol{I}(2h) = \boldsymbol{P}_1(h)\boldsymbol{P}_1(h)\boldsymbol{I}(0). \tag{23b}$$

Since

$$\boldsymbol{I}(2h) = \boldsymbol{P}_2(2h)\boldsymbol{I}(0),$$

Eq(23b) specifies

$$\boldsymbol{P}_2(2h) \equiv \boldsymbol{P}_1(h)^2. \tag{24}$$

From Eq(11), for the single slab

$$\boldsymbol{P}_1(h) = \begin{bmatrix} \boldsymbol{T}^{-1}(h) & -\boldsymbol{T}^{-1}(h)\boldsymbol{R}(h) \\ \boldsymbol{R}(h)\boldsymbol{T}^{-1}(h) & \boldsymbol{T}(h) - \boldsymbol{R}(h)\boldsymbol{T}^{-1}(h)\boldsymbol{R}(h) \end{bmatrix}, \tag{25a}$$

and from Eqs (11), (23c) and (24)

$$\begin{aligned} \boldsymbol{P}_2(2h) &\equiv \begin{bmatrix} \boldsymbol{T}^{-1}(2h) & -\boldsymbol{T}^{-1}(2h)\boldsymbol{R}(2h) \\ \boldsymbol{R}(2h)\boldsymbol{T}^{-1}(2h) & \boldsymbol{T}(2h) - \boldsymbol{R}(2h)\boldsymbol{T}^{-1}(2h)\boldsymbol{R}(2h) \end{bmatrix} \\ &= \begin{bmatrix} \boldsymbol{T}^{-1}(h) & -\boldsymbol{T}^{-1}(h)\boldsymbol{R}(h) \\ \boldsymbol{R}(h)\boldsymbol{T}^{-1}(h) & \boldsymbol{T}(h) - \boldsymbol{R}(h)\boldsymbol{T}^{-1}(h)\boldsymbol{R}(h) \end{bmatrix}^2. \end{aligned} \tag{25b}$$

Equating the four matrix partitions across the equal sign results in the following well-known propagated interaction coefficients for a two identical slab composite [11]:

$$\begin{aligned} \boldsymbol{T}_2 &= \boldsymbol{T}_1\left[\boldsymbol{I}_N - \boldsymbol{R}_1^2\right]^{-1}\boldsymbol{T}_1 \\ \boldsymbol{R}_2 &= \boldsymbol{R}_1 + \boldsymbol{T}_1\boldsymbol{R}_1\left[\boldsymbol{I}_N - \boldsymbol{R}_1^2\right]^{-1}\boldsymbol{T}_1, \end{aligned} \tag{26a}$$

where

$$\begin{aligned} \boldsymbol{T}_1 &\equiv \boldsymbol{T}(h) \\ \boldsymbol{T}_2 &\equiv \boldsymbol{T}(2h) \\ \boldsymbol{R}_1 &\equiv \boldsymbol{R}(h) \\ \boldsymbol{R}_2 &\equiv \boldsymbol{R}(2h). \end{aligned} \tag{26b}$$

The identical relation should also be true from Eq(20b). With the determination of the interaction coefficients, the corresponding composite response for two identical slabs follows as

$$\boldsymbol{Q}_2 \equiv \begin{bmatrix} \boldsymbol{T}_2 & \boldsymbol{R}_2 \\ \boldsymbol{R}_2 & \boldsymbol{T}_2 \end{bmatrix}. \tag{26c}$$

We next double the two-slab response to construct a four-slab response and then an eight-slab response etc., and continue until we construct the response matrix for the entire slab of thickness $\tau$. We achieve this by initially dividing the slab into $2^n$ slabs each of thickness

$$h_n \equiv \frac{\tau}{2^n}. \tag{27a}$$

Doubling to thickness $2^l h_n$, for $l$ =1,…,$n$, the interaction coefficients from Eq(26a) are

$$\begin{aligned} \boldsymbol{T}_{2^l} &= \boldsymbol{T}_{2^{l-1}}\left[\boldsymbol{I}_N - \boldsymbol{R}_{2^{l-1}}^2\right]^{-1}\boldsymbol{T}_{2^{l-1}} \\ \boldsymbol{R}_{2^l} &= \boldsymbol{R}_{2^{l-1}} + \boldsymbol{T}_{2^{l-1}}\boldsymbol{R}_{2^{l-1}}\left[\boldsymbol{I}_N - \boldsymbol{R}_{2^{l-1}}^2\right]^{-1}\boldsymbol{T}_{2^{l-1}} \end{aligned} \tag{27b,c}$$

with coverage when $l = n$

$$\begin{bmatrix} \boldsymbol{I}_0^- \\ \boldsymbol{I}_{2^n}^+ \end{bmatrix} = \begin{bmatrix} \boldsymbol{T}_{2^n} & \boldsymbol{R}_{2^n} \\ \boldsymbol{R}_{2^n} & \boldsymbol{T}_{2^n} \end{bmatrix}\begin{bmatrix} \boldsymbol{I}_{2^n}^- \\ \boldsymbol{I}_0^+ \end{bmatrix} = \boldsymbol{Q}_{2^n}\begin{bmatrix} \boldsymbol{I}_{2^n}^- \\ \boldsymbol{I}_0^+ \end{bmatrix}, \tag{28a}$$

or

$$\begin{bmatrix} \boldsymbol{I}^-(0) \\ \boldsymbol{I}^+(\tau) \end{bmatrix} = \begin{bmatrix} \boldsymbol{T}(\tau) & \boldsymbol{R}(\tau) \\ \boldsymbol{R}(\tau) & \boldsymbol{T}(\tau) \end{bmatrix}\begin{bmatrix} \boldsymbol{I}^-(\tau) \\ \boldsymbol{I}^+(0) \end{bmatrix} = \boldsymbol{Q}(\tau)\begin{bmatrix} \boldsymbol{I}^-(\tau) \\ \boldsymbol{I}^+(0) \end{bmatrix}. \tag{28b}$$

The significant numerical advantage of doubling is immediately apparent since rather than marching discretely from the near to the far surface in $2^n$ steps, one doubles to the far surface in $n$ steps greatly reducing the propagation error.

Note that after Eq(25b) $\boldsymbol{P}(\tau)$ is not required in this formulation, though in other formulations it is [10].

### 3.2 Interior Intensity by Doubling

There are several choices to find the interior intensity. The first and least advantageous is from the interaction principle itself. From Eq(5b)

$$\boldsymbol{I}(\tau) = \boldsymbol{P}(\tau)\boldsymbol{I}(0)$$

when applied recursively over $l$ doublings to $\tau = 2^l h_l$ to give

$$\boldsymbol{I}(\tau) = \boldsymbol{P}_{2^l}\left(2^l h_l\right)\boldsymbol{I}(0), \tag{29a}$$

where $\boldsymbol{P}_{2^l}$ is constructed from

$$\boldsymbol{P}_{2^j}\left(2^j h_l\right) \equiv \begin{bmatrix} \boldsymbol{T}_{2^j}^{-1}\left(2^j h_l\right) & -\boldsymbol{T}_{2^j}^{-1}\left(2^j h_l\right)\boldsymbol{R}_{2^j}\left(2^j h_l\right) \\ \boldsymbol{R}_{2^j}\left(2^j h_l\right)\boldsymbol{T}_{2^j}^{-1}\left(2^j h_l\right) & \boldsymbol{T}_{2^j}\left(2^j h_l\right) - \boldsymbol{R}_{2^j}\left(2^j h_l\right)\boldsymbol{T}_{2^j}^{-1}\left(2^j h_l\right)\boldsymbol{R}_{2^j}\left(2^j h_l\right) \end{bmatrix} \tag{29b}$$

for $j = 1,\ldots l$. Thus, the interior intensity distribution is

$$\begin{bmatrix} \boldsymbol{I}_l^- \\ \boldsymbol{I}_l^+ \end{bmatrix} = \begin{bmatrix} \boldsymbol{P}_{l,1} & \boldsymbol{P}_{l,2} \\ \boldsymbol{P}_{l,3} & \boldsymbol{P}_{l,4} \end{bmatrix} \begin{bmatrix} \boldsymbol{I}_0^- \\ \boldsymbol{I}_0^+ \end{bmatrix}. \tag{29c}$$

The disadvantage here is that the exiting intensity leaving the near face $\boldsymbol{I}_0^-$ must be determined first. In addition, the matrix inversion of the ill-conditioned transmission coefficient is required.

More straightforwardly, from Eq(21) becomes, as referenced to Fig. 4

$$\begin{bmatrix} \boldsymbol{I}^+(\tau) \\ \boldsymbol{I}^-(\tau) \end{bmatrix} = \boldsymbol{U}(\tau, \tau_L - \tau) \begin{bmatrix} \boldsymbol{I}_L^- \\ \boldsymbol{I}_0^+ \end{bmatrix}. \tag{30}$$

Now only the prescribed incoming intensities are required without any matrix inversion. To apply Eq(30), one needs to find $\boldsymbol{Q}(\tau)$, $\boldsymbol{R}(\tau_L - \tau)$ by doubling and $\boldsymbol{U}(\tau, \tau_L - \tau)$ from Eqs(20e).

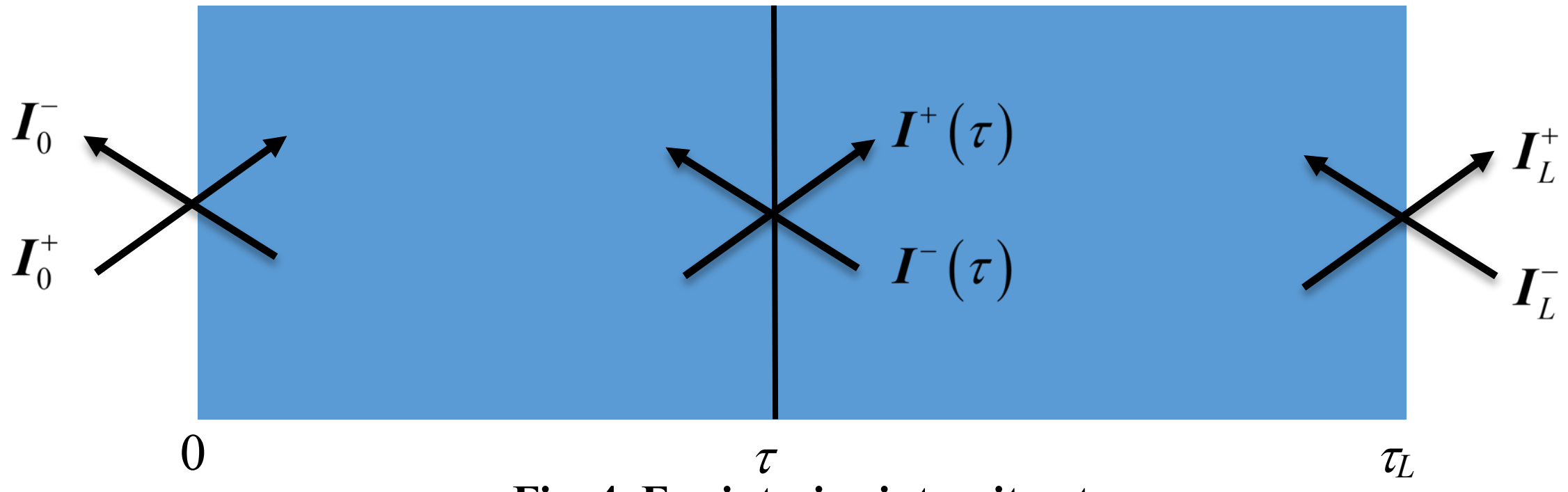

**Fig. 4. For interior intensity at $\tau$.**

### 3.3 Interaction Coefficients by Taylor series

The fundamental numerical algorithm depends upon the evaluation of the interaction coefficients of reflectance and transmission by TS

$$\boldsymbol{F}(\tau) = \sum_{n=0}^{\infty} \boldsymbol{F}_n h^n. \tag{31}$$

$\boldsymbol{F}$ is either of the quantities $\boldsymbol{R}$ or $\boldsymbol{T}$. The coefficients $\boldsymbol{F}_n$ are determined recursively for any $n$ as shown above and in APPENDIX B. Each TS is for small slab of width $h$ to begin doubling as discussed below and nothing more. For this reason, we expect rapid convergence. The series sum continues until the last value of all components of the matrix $\boldsymbol{F}_n$ are within a prescribed relative error to the total sum.

### 3.4 Additional Analytical Considerations

Several relatively unfamiliar analytical features aid in the implementation of MREM, including faux quadrature interpolation, exact beam source and the grazing angle.

#### 3.4.1 Faux quadrature interpolation

As will be shown, a key element of the benchmark from MREM will be faux (false) quadrature interpolation. Interpolation based on the matrix Riccati equations themselves, proceeds as follows. One can add additional (faux) quadrature abscissae to the quadrature list provided they have zero weight. Thus, the solution algorithm gives interpolated intensities at the faux abscissae and their addition does not modify the solution in any way. Faux quadrature can highlight a particular direction range or insure a particular set of directions is in each grid approximation for error assessment.

#### 3.4.2 Exact beam source

Generally, beam source implementation in discrete ordinates methods is thought to be limited by the quadrature set chosen. However, a beam along a specific direction is easily implemented using faux quadrature when one includes the beam direction in the quadrature list and sets the beam intensity to $S_0/\varepsilon$ in that direction and zero for all other directions. The beam quadrature weight is then set to $\varepsilon$ so that integration over all directions normalizes the beam source to $S_0$. Thus, the source is included in its actual direction. In this way, we have avoided the need to do a separate determination of the uncollided contribution. The intensity approximation is insensitive to the value of $\varepsilon$ as long as it is small ($\varepsilon < 10^{-16}$).

#### 3.4.3 Intensity at $\mu = 0$

The chosen quadrature cannot accommodate $\mu = 0$ either directly or through faux quadrature interpolation. Thus, $\mu = 0$ requires special treatment. In particular, from Eq(1a) with $\mu_m = 0$, we have (with argument $N$ suppressed)

$$I(\tau,0)=\sum_{m'=1}^{2N}\alpha_{m'}f(0,\mu_{m'})I(\tau,\mu_{m'}),\ 0<\tau<\tau_1. \tag{32a}$$

or in terms of the directional intensities at $\tau$

$$I(\tau,0)=\sum_{m'=1}^{N}\alpha_{m'}\left[f(0,\mu_{m'})I_{m'}^{+}(\tau)+f(0,\mu_{N+m'})I_{m'}^{-}(\tau)\right],\ 0<\tau<\tau_1. \tag{32b}$$

Since $I^{\pm}(\tau)$ comes from Eq(30) one can evaluate this expression. In the limit as $\tau$ approaches the two surfaces

$$I(0,0)=\sum_{m'=1}^{N}\alpha_{m'}\left[f(0,\mu_{m'})g_{m'}(0)+f(0,\mu_{N+m'})I^{-}_{m'}(0)\right]$$
$$I(0,\tau)=\sum_{m'=1}^{N}\alpha_{m'}f(0,\mu_{m'})I^{+}_{m'}(\tau), \quad (32c,d)$$

where the required exiting intensities $\boldsymbol{I}^{-}(0),\boldsymbol{I}^{+}(\tau)$ come from Eq(28b).

## 4. NUMERICAL IMPLEMENTATION

This section considers the additional numerical features to make the MREM a high precision method for up to seven digits of the true solution.

### 4.1 Wynn-Epsilon Convergence Acceleration

The Wynn-epsilon (*W-e*) convergence algorithm [12] plays a significant role in achieving high precision. The concept of convergence acceleration is to ensure the limit of a sequence of solutions, for example based on number of discrete directions *N*, converges to the true solution. Convergence however, does not need be sequential with increasing number of discretizations *N*. If we know the asymptotic behavior of the sequence, then a surrogate sequence can be established to possibly converge to the limit more quickly. The Wynn-epsilon algorithm is one such algorithm based on a non-linear recurrence to find the asymptotic trend in a sequence for a large variety of sequences, however not all.

#### 4.1.1 Acceleration for interaction coefficients

Built on directional discretization and TS, MREM offers several opportunities to apply convergence acceleration beginning with the TS. The TS representation of Eq(31) written as the limit of partial sums is

$$\boldsymbol{F}(\tau)=\lim_{m\to\infty}\sum_{n=0}^{m}\boldsymbol{F}_n\tau^n. \quad (33)$$

The partial sums therefore form a sequence to which one can apply the following non-linear *W-e* recurrence by component

$$\begin{aligned}&\varepsilon_{-1}^{(m)}\equiv 0\\&\varepsilon_{0}^{(m)}\equiv f_{m,i,j},\ m=0,\dots,L\\&\varepsilon_{k+1}^{(m)}=\varepsilon_{k-1}^{(m+1)}+\left[\varepsilon_{k}^{(m+1)}-\varepsilon_{k}^{(m)}\right]^{-1},\ k=0,\dots,2K-1\ ;\ m=0,\dots,L,\end{aligned} \quad (34)$$

where $f_{m,i,j}$ is the sequence of components of the matrix sequence $\hat{\boldsymbol{F}}_m$ of partial sums

$$\hat{\boldsymbol{F}}_m(\tau) \equiv \sum_{n=0}^{m} \boldsymbol{F}_n \tau^n. \tag{35}$$

for $m = 0,\ldots$ . Each $\varepsilon_{k+1}^{(m)}$ represents a generally more quickly convergent approximation to the limit. These quantities form the *W-e* Tableau shown in Fig. 5, with every other term on the last diagonal taken to be a more quickly convergent sequence than the original.

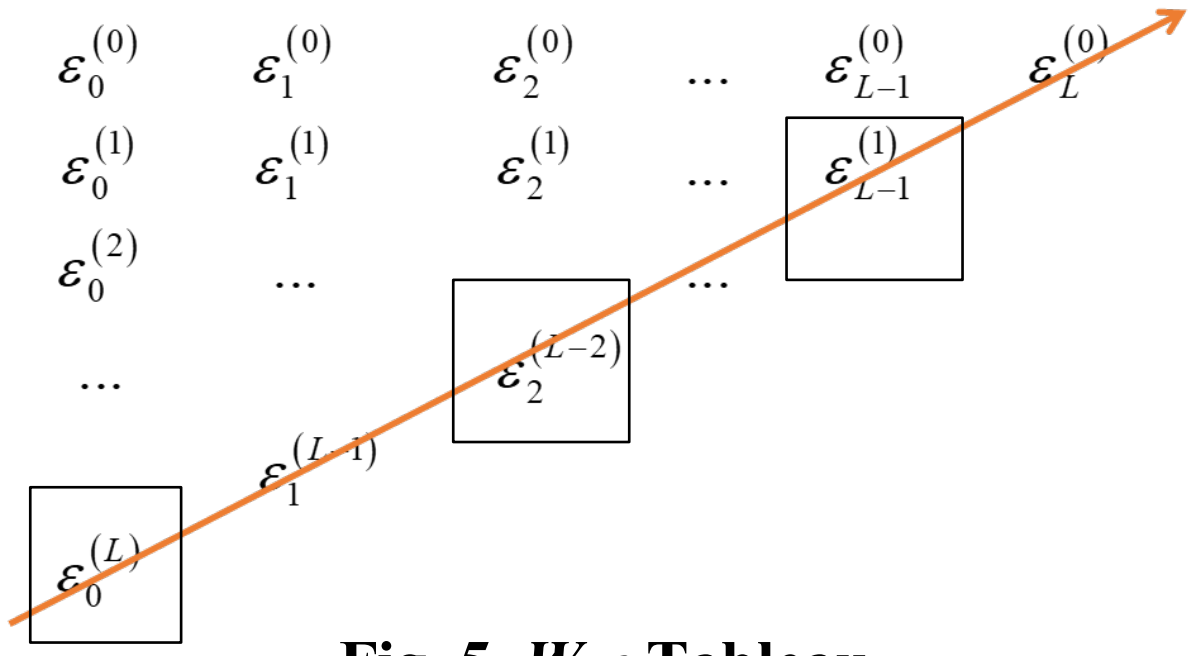

**Fig. 5. *W-e* Tableau.**

As an example, Table I shows the advantage of convergence acceleration for the TS

**Table I: TS Relative Errors for fixed *h* and *N*.**

| *m* | *R*(*Ori*) | *Ratio* | *T*(*Ori*) | *Ratio* |
|---|---|---|---|---|
| 2 | 9.48E-01 | 1.00E+00 | 9.48E-01 | 1.00E+00 |
| 3 | 1.39E-05 | 1.00E+00 | 1.36E-05 | 1.00E+00 |
| 4 | 2.91E-07 | 1.02E+00 | 2.81E-07 | 1.02E+00 |
| 5 | 5.03E-09 | 3.97E+00 | 4.74E-09 | 3.94E+00 |
| 6 | 7.37E-11 | 4.89E+00 | 6.64E-11 | 4.85E+00 |
| 7 | 9.49E-13 | 5.29E+00 | 7.84E-13 | 5.12E+00 |

evaluation of ***R*** and ***T*** at the slab surface for the Cloud C1 phase function considered in the next section. The table shows the relative errors for a given *h* used in doubling and quadrature order *N* and their ratio for the original (***Ori***) and *W-e* sequence (not given). In our context, converged means the difference between two TS approximations is less than a desired amount. One observes the ratio (***Ratio***) is generally larger than unity indicating superiority of *W-e* convergence.

Since the interaction coefficients are at the center of the calculation, their efficient determination will significantly reduce computational time. For this reason, the number of terms in the TS is limited (~15). If not converged, we then take the next *h*, as explained in the next section.

### 4.1.2 Acceleration of doubling

A second example of convergence acceleration, though less dramatic, comes from the choice of *h* to determine the slab interaction coefficients leading to intensities $I_m^{\pm}$ from doubling. As shown in Ref. [4], the determination of *h* to start doubling is nontrivial. We will exploit some known trends that enable precise determination. In particular, if *h* is too large, the reflectance and transmittance become unphysically negative. To capture this, we choose an initial *h* and check for positivity. If the check fails, we replace *h* by *h*/2 and check again until positive. Positive interaction coefficients marks the beginning of the sequence choice $h_{l_2}$ for slab thickness $\tau$

$$h_l \equiv \frac{\tau}{2^l};\ l = l_1,\ldots,l_2\,. \tag{36a}$$

It is generally found that after initial positivity, the coefficients remain positive for smaller *h*. Thus, in this way we develop a sequence of solutions in *h* for a fixed quadrature order. One also need be aware of round off error, which destroys the calculation if $h_l$ becomes too small. Thus, there is an optimum *h*. Care must be taken to treat this delicate balance. One accomplishes this by assessing both the original (*Ori*) and *W-e* accelerated relative errors between consecutive *h*-iterations

$$\begin{aligned} Rel^{\pm}_{Ori,l} &\equiv \left|I^{\pm}_m(\tau;h_l) - I^{\pm}_m(\tau;h_{l-1})\right| / \left.I^{\pm}_m(\tau;h_l)\right|_{Ori} \\ Rel^{\pm}_{We,l} &\equiv \left|I^{\pm}_m(\tau;h_l) - I^{\pm}_m(\tau;h_{l-1})\right| / \left.I^{\pm}_m(\tau;h_l)\right|_{We}. \end{aligned} \tag{36b}$$

where $h_l$ is now included in the argument. The converged solution is the one satisfying

$$\min_{l_1 \le l \le l_2}\left[Rel^{\pm}_{Ori,l}, Rel^{\pm}_{We,l}\right]. \tag{36c}$$

Since this condition may not meet the desired relative error, there is a minimum relative error achievable with the MREM found to be $10^{-8}$.

Table II shows the corruption of relative error by round off error with increasing *l* for the

**Table II: Relative error sequence for choice of *h***

| *l* | *R*(*Ori*) | *T*(*Ori*) | *R*(*W-e*) | *T*(*W-e*) |
|---|---|---|---|---|
| 23 | 1.09E+00 | 1.09E+00 | 1.09E+00 | 1.09E+00 |
| 24 | 3.57E-09 | 1.33E-08 | 3.57E-09 | 1.33E-08 |
| 25 | 1.07E-08 | 4.01E-08 | 5.35E-09 | 2.00E-08 |
| 26 | 3.19E-08 | 1.20E-07 | 7.37E-11 | 7.93E-11 |
| 27 | 6.51E-08 | 2.46E-07 | 1.45E-08 | 5.57E-08 |
| 28 | 7.12E-08 | 2.66E-07 | 6.52E-07 | 2.77E-06 |
| 29 | 6.50E-08 | 2.41E-07 | 1.59E-06 | 6.08E-06 |
| 30 | 8.32E-08 | 3.11E-07 | 9.75E-07 | 3.40E-06 |

Cloud C1 phase function. The errors shown are averaged over 22 faux interpolation edits. The table values indicate that *W-e* for *l* = 26 give the minimal relative error shown shaded. Note that *l* = 23 is the first solution with all positive interaction coefficients.

### 4.1.3 Acceleration of the quadrature approximation

Recall the solution to the radiative transfer equation is the limit

$$I_m^{\pm}(\tau) \equiv \lim_{N\to\infty} I(\tau, \pm\mu_m; N), \tag{37}$$

which is precisely where *W-e* would most appropriately apply. For a series of discrete *N*, $I(\tau, \pm\mu_m; N)$ forms a sequence of solutions in *N* converging to the true solution as indicated. If for each $\tau$ and $\mu_m$, one introduces the sequence into the *W-e* algorithm [Eq(34)] the approximate solution will most likely be accelerated towards its true solution. Since convergence acceleration is an experimental mathematical procedure and experiments are not always successful, *W-e* will not always better convergence that the original sequence.

For the Cloud C1 phase function, convergence of the quadrature sequence based on relative error between consecutive approximations is equally effective with or without convergence acceleration. Therefore the inclusion of *W-e* is more for completeness and answers the question of whether it is effective of not. Otherwise, the question would remain unanswered.

Since directions change with each Gauss quadrature order, faux abscissae, which remain invariant with order, provide the measure of convergence of the angular intensity. Convergence is based on the relative error of the faux quadrature set between consecutive orders, which are given a stride of 10.

## 5. BENCHMARKS

We now come to benchmarks-- the focus of this work. The benchmark format follows that of Ref. [6]. We consider two standard benchmarks—the HAZE L and Cloud C1 phase functions [13], representing the state of art in radiative transfer verification. APPENDIX C gives both phase functions for completeness. With the beam source normalized to ½, the tables of intensities presented are precise to a solid six places on rounding with many to seven places. The error measure is for uniform faux quadrature abscissae in the interval [-1,1]. Both benchmarks agree to all five places of Siewert's pioneering ADO benchmarks [6]. The first author's response matrix method (RM/DOM) will serve as the standard since RM/DOM should be precise to one unit in the 7$^{th}$ place.

To the authors' knowledge, no other benchmarks exist to the precision presented here.

### 5.1. HAZE L Phase Function

The first benchmark is for a HAZE L conservative scattering atmosphere of one *mfp* thickness. The source is a perpendicular beam on the near surface ($\tau = 0$). APPENDIX C gives the 82 term scattering kernel. Table IIIa shows intensities uniformly distributed in the direction interval [-1,1] correct to one unit in the last place indicated by an emboldened digit on comparison with Ref 10. Table IIIb shows six places when compared to Table IIIa.

**Table IIIa: HAZE L /~ Seven-places/*N*~190/*h* =$2^{-23}$/~150*s*/**

| $\mu\backslash\tau$ | 0 | $\tau_0/20$ | $\tau_0/10$ | $\tau_0/5$ | $\tau_0/2$ | $3\tau_0/4$ | $\tau_0$ |
|---|---|---|---|---|---|---|---|
| -1.000E+00 | 3.6145155E-02 | 3.4339396E-02 | 3.2510866E-02 | 2.8812216E-02 | 1.7628611E-02 | 8.5258907E-03 | 0.0000000E-00 |
| -9.000E-01 | 3.9781870E-02 | 3.7872320E-02 | 3.5920681E-02 | 3.1930313E-02 | 1.9620173E-02 | 9.4573132E-03 | 0.0000000E-00 |
| -8.000E-01 | 4.2731263E-02 | 4.0840607E-02 | 3.8873442E-02 | 3.4767734E-02 | 2.1601856E-02 | 1.0395856E-02 | 0.0000000E-00 |
| -7.000E-01 | 4.8005146E-02 | 4.6131929E-02 | 4.4130697E-02 | 3.9829198E-02 | 2.5247889E-02 | 1.2217079E-02 | 0.0000000E-00 |
| -6.000E-01 | 5.5821352E-02 | 5.4043176E-02 | 5.2059351E-02 | 4.7598582E-02 | 3.1183659E-02 | 1.5361836E-02 | 0.0000000E-00 |
| -5.000E-01 | 6.6094220E-02 | 6.4629636E-02 | 6.2844874E-02 | 5.8497071E-02 | 4.0273976E-02 | 2.0562126E-02 | 0.0000000E-00 |
| -4.000E-01 | 7.8148081E-02 | 7.7440255E-02 | 7.6250768E-02 | 7.2704872E-02 | 5.3729974E-02 | 2.9128534E-02 | 0.0000000E-00 |
| -3.000E-01 | 8.9968154E-02 | 9.0770641E-02 | 9.0878384E-02 | 8.9471128E-02 | 7.2964349E-02 | 4.3468799E-02 | 0.0000000E-00 |
| -2.000E-01 | 9.7081540E-02 | 1.0042085E-01 | 1.0278927E-01 | 1.0550594E-01 | 9.8377714E-02 | 6.7994924E-02 | 0.0000000E-00 |
| -1.000E-01 | 9.2932814E-02 | 9.9818714E-02 | 1.0519502E-01 | 1.1349749E-01 | 1.2403692E-01 | 1.0839912E-01 | 0.0000000E-00 |
| 0.000E+00 | 6.9877391E-02 | 8.4667310E-02 | 9.4166298E-02 | 1.0872694E-01 | 1.3576248E-01 | 1.4277947E-01 | 0.0000000E-00 |
| 0.000E+00 | 0.0000000E-00 | 8.4667310E-02 | 9.4166298E-02 | 1.0872694E-01 | 1.3576248E-01 | 1.4277947E-01 | 1.1480771E-01 |
| 1.000E-01 | 0.0000000E-00 | 2.9541820E-02 | 5.2434564E-02 | 8.4564914E-02 | 1.3509602E-01 | 1.5610649E-01 | 1.5697621E-01 |
| 2.000E-01 | 0.0000000E-00 | 1.6490680E-02 | 3.2281653E-02 | 6.0752687E-02 | 1.2435036E-01 | 1.5892546E-01 | 1.7681766E-01 |
| 3.000E-01 | 0.0000000E-00 | 1.2342100E-02 | 2.4848764E-02 | 4.9396789E-02 | 1.1481121E-01 | 1.5793652E-01 | 1.8830112E-01 |
| 4.000E-01 | 0.0000000E-00 | 1.1187938E-02 | 2.2644990E-02 | 4.5754673E-02 | 1.1226864E-01 | 1.6086203E-01 | 2.0001869E-01 |
| 5.000E-01 | 0.0000000E-00 | 1.1795942E-02 | 2.3790964E-02 | 4.8000306E-02 | 1.1907946E-01 | 1.7319074E-01 | 2.1963289E-01 |
| 6.000E-01 | 0.0000000E-00 | 1.4204907E-02 | 2.8458379E-02 | 5.6873102E-02 | 1.3905102E-01 | 2.0144487E-01 | 2.5598334E-01 |
| 7.000E-01 | 0.0000000E-00 | 1.9583294E-02 | 3.8924847E-02 | 7.6745368E-02 | 1.8200357E-01 | 2.5898644E-01 | 3.2512495E-01 |
| 8.000E-01 | 0.0000000E-00 | 3.1953231E-02 | 6.2942983E-02 | 1.2204484E-01 | 2.7718191E-01 | 3.8276705E-01 | 4.6865778E-01 |
| 9.000E-01 | 0.0000000E-00 | 6.8726703E-02 | 1.3391677E-01 | 2.5425935E-01 | 5.4460066E-01 | 7.1944669E-01 | 8.4608373E-01 |
| 1.000E+00 | 0.0000000E-00 | 3.6493954E-01 | 7.0026634E-01 | 1.2895497E+00 | 2.5225517E+00 | 3.0931861E+00 | 3.3809098E+00 |

**Table IIIb: HAZE L/Six-places/*N*~190/*h* =$2^{-23}$/~60*s*/**

| $\mu\backslash\tau$ | 0 | $\tau_0/20$ | $\tau_0/10$ | $\tau_0/5$ | $\tau_0/2$ | $3\tau_0/4$ | $\tau_0$ |
|---|---|---|---|---|---|---|---|
| -1.000E+00 | 3.614516E-02 | 3.433940E-02 | 3.251087E-02 | 2.881222E-02 | 1.762861E-02 | 8.525891E-03 | 0.000000E-00 |
| -9.000E-01 | 3.978187E-02 | 3.787232E-02 | 3.592068E-02 | 3.193031E-02 | 1.962017E-02 | 9.457313E-03 | 0.000000E-00 |
| -8.000E-01 | 4.273126E-02 | 4.084061E-02 | 3.887344E-02 | 3.476773E-02 | 2.160186E-02 | 1.039586E-02 | 0.000000E-00 |
| -7.000E-01 | 4.800515E-02 | 4.613193E-02 | 4.413070E-02 | 3.982920E-02 | 2.524789E-02 | 1.221708E-02 | 0.000000E-00 |
| -6.000E-01 | 5.582135E-02 | 5.404318E-02 | 5.205935E-02 | 4.759858E-02 | 3.118366E-02 | 1.536184E-02 | 0.000000E-00 |
| -5.000E-01 | 6.609422E-02 | 6.462964E-02 | 6.284487E-02 | 5.849707E-02 | 4.027398E-02 | 2.056213E-02 | 0.000000E-00 |
| -4.000E-01 | 7.814808E-02 | 7.744025E-02 | 7.625077E-02 | 7.270487E-02 | 5.372997E-02 | 2.912853E-02 | 0.000000E-00 |
| -3.000E-01 | 8.996815E-02 | 9.077064E-02 | 9.087838E-02 | 8.947113E-02 | 7.296435E-02 | 4.346880E-02 | 0.000000E-00 |
| -2.000E-01 | 9.708154E-02 | 1.004209E-01 | 1.027893E-01 | 1.055059E-01 | 9.837771E-02 | 6.799492E-02 | 0.000000E-00 |
| -1.000E-01 | 9.293281E-02 | 9.981871E-02 | 1.051950E-01 | 1.134975E-01 | 1.240369E-01 | 1.083991E-01 | 0.000000E-00 |
| 0.000E+00 | 6.987739E-02 | 8.466731E-02 | 9.416630E-02 | 1.087269E-01 | 1.357625E-01 | 1.427795E-01 | 0.000000E-00 |
| 0.000E+00 | 0.000000E-00 | 8.466731E-02 | 9.416630E-02 | 1.087269E-01 | 1.357625E-01 | 1.427795E-01 | 1.148077E-01 |
| 1.000E-01 | 0.000000E-00 | 2.954182E-02 | 5.243456E-02 | 8.456491E-02 | 1.350960E-01 | 1.561065E-01 | 1.569762E-01 |
| 2.000E-01 | 0.000000E-00 | 1.649068E-02 | 3.228165E-02 | 6.075269E-02 | 1.243504E-01 | 1.589255E-01 | 1.768177E-01 |
| 3.000E-01 | 0.000000E-00 | 1.234210E-02 | 2.484876E-02 | 4.939679E-02 | 1.148112E-01 | 1.579365E-01 | 1.883011E-01 |
| 4.000E-01 | 0.000000E-00 | 1.118794E-02 | 2.264499E-02 | 4.575467E-02 | 1.122686E-01 | 1.608620E-01 | 2.000187E-01 |
| 5.000E-01 | 0.000000E-00 | 1.179594E-02 | 2.379096E-02 | 4.800031E-02 | 1.190795E-01 | 1.731907E-01 | 2.196329E-01 |
| 6.000E-01 | 0.000000E-00 | 1.420491E-02 | 2.845838E-02 | 5.687310E-02 | 1.390510E-01 | 2.014449E-01 | 2.559833E-01 |
| 7.000E-01 | 0.000000E-00 | 1.958329E-02 | 3.892485E-02 | 7.674537E-02 | 1.820036E-01 | 2.589864E-01 | 3.251249E-01 |
| 8.000E-01 | 0.000000E-00 | 3.195323E-02 | 6.294298E-02 | 1.220448E-01 | 2.771819E-01 | 3.827671E-01 | 4.686578E-01 |
| 9.000E-01 | 0.000000E-00 | 6.872670E-02 | 1.339168E-01 | 2.542594E-01 | 5.446007E-01 | 7.194467E-01 | 8.460837E-01 |
| 1.000E+00 | 0.000000E-00 | 3.649395E-01 | 7.002663E-01 | 1.289550E+00 | 2.522552E+00 | 3.093186E+00 | 3.380910E+00 |

For doubling, *l* varied form $l_1 = 21$ to $l_2 = 25$ and the quadrature order was about $N = 190$. From experimentation, it seems that MREM is more appropriate for small slabs; while, the RM/DOM is best for larger slabs. The computational time for these table is 150*s* and 60*s* respectively.

## 5.2 Cloud C1 Phase Function

The second benchmark is for the Cloud C1 conservative scattering atmosphere of 64 *mfp* thickness. Again, the source is a perpendicular beam on the near surface ($\tau = 0$). The 300 term scattering kernel is given in APPENDIX C and represents a true challenge. Table IVa

**Table IVa: Cloud C1//~ Seven-places /*N*~190/*h* =$2^{-23}$/~265*s*/**

| $\mu \backslash \tau$ | 0 | $\tau_0/20$ | $\tau_0/10$ | $\tau_0/5$ | $\tau_0/2$ | $3\tau_0/4$ | $\tau_0$ |
|---|---|---|---|---|---|---|---|
| -1.000E+00 | 1.0636984E+00 | 1.0062387E+00 | 9.6320640E-01 | 8.582423**2**E-01 | 5.2453336E-01 | 2.4600228E-01 | 0.0000000E-00 |
| -9.000E-01 | 9.5309008E-01 | 9.956622**8**E-01 | 9.697242**0**E-01 | 8.693898**2**E-01 | 5.3598880E-01 | 2.5740482E-01 | 0.0000000E-00 |
| -8.000E-01 | 9.5407647E-01 | 9.9828274E-01 | 9.777659**0**E-01 | 8.805282**2**E-01 | 5.474442**6**E-01 | 2.6883574E-01 | 0.0000000E-00 |
| -7.000E-01 | 8.8254184E-01 | 9.885061**3**E-01 | 9.8351863E-01 | 8.915683**4**E-01 | 5.5889973E-01 | 2.8028148E-01 | 0.0000000E-00 |
| -6.000E-01 | 8.2471232E-01 | 9.7909867E-01 | 9.889035**9**E-01 | 9.025562**9**E-01 | 5.703551**7**E-01 | 2.9173427E-01 | 0.0000000E-00 |
| -5.000E-01 | 7.7260568E-01 | 9.697724**0**E-01 | 9.939905**5**E-01 | 9.1349752E-01 | 5.818106**0**E-01 | 3.0319021E-01 | 0.0000000E-00 |
| -4.000E-01 | 7.1143850E-01 | 9.5800446E-01 | 9.983295**5**E-01 | 9.2437587E-01 | 5.9326601E-01 | 3.1464747E-01 | 0.0000000E-00 |
| -3.000E-01 | 6.4031056E-01 | 9.4342529E-01 | 1.0017757E+00 | 9.351789**6**E-01 | 6.0472139E-01 | 3.2610518E-01 | 0.0000000E-00 |
| -2.000E-01 | 5.5848173E-01 | 9.2583435E-01 | 1.0042140E+00 | 9.458945**0**E-01 | 6.161767**3**E-01 | 3.3756297E-01 | 0.0000000E-00 |
| -1.000E-01 | 4.5873404E-01 | 9.045925**0**E-01 | 1.0054770E+00 | 9.5650947E-01 | 6.2763204E-01 | 3.4902064E-01 | 0.0000000E-00 |
| 0.000E+00 | 2.5158245E-01 | 8.795199**8**E-01 | 1.0054752E+00 | 9.670143**4**E-01 | 6.3908730E-01 | 3.6047812E-01 | 0.0000000E-00 |
| 0.000E+00 | 0.0000000E-00 | 8.795199**8**E-01 | 1.0054752E+00 | 9.670143**4**E-01 | 6.3908730E-01 | 3.6047812E-01 | 0.0000000E-00 |
| 1.000E-01 | 0.0000000E-00 | 8.5069625E-01 | 1.0042130E+00 | 9.774071**8**E-01 | 6.5054252E-01 | 3.7193540E-01 | 7.2039069E-02 |
| 2.000E-01 | 0.0000000E-00 | 8.1871840E-01 | 1.0018813E+00 | 9.877015**2**E-01 | 6.619976**9**E-01 | 3.8339248E-01 | 8.8948974E-02 |
| 3.000E-01 | 0.0000000E-00 | 7.8521516E-01 | 9.990155**2**E-01 | 9.979409**3**E-01 | 6.7345285E-01 | 3.9484937E-01 | 1.0414040E-01 |
| 4.000E-01 | 0.0000000E-00 | 7.545985**4**E-01 | 9.9678837E-01 | 1.0082267E+00 | 6.8490802E-01 | 4.0630611E-01 | 1.1838392E-01 |
| 5.000E-01 | 0.0000000E-00 | 7.351621**3**E-01 | 9.976028**2**E-01 | 1.0187710E+00 | 6.9636329E-01 | 4.1776272E-01 | 1.3199043E-01 |
| 6.000E-01 | 0.0000000E-00 | 7.376549**3**E-01 | 1.0059571E+00 | 1.030001**6**E+00 | 7.078188**3**E-01 | 4.2921921E-01 | 1.4513931E-01 |
| 7.000E-01 | 0.0000000E-00 | 7.779240**6**E-01 | 1.0300617E+00 | 1.042781**8**E+00 | 7.1927503E-01 | 4.4067560E-01 | 1.5795774E-01 |
| 8.000E-01 | 0.0000000E-00 | 8.870455**2**E-01 | 1.0859454E+00 | 1.058910**5**E+00 | 7.3073267E-01 | 4.5213189E-01 | 1.7054334E-01 |
| 9.000E-01 | 0.0000000E-00 | 1.1539390E+00 | 1.214126**8**E+00 | 1.082650**3**E+00 | 7.421935**8**E-01 | 4.6358809E-01 | 1.8295432E-01 |
| 1.000E+00 | 0.0000000E-00 | 8.074596**4**E+01 | 1.1786122E+01 | 1.260595**9**E+00 | 7.536634**9**E-01 | 4.7504419E-01 | 1.9523120E-01 |

**Table IVb: Cloud C1//Six-places /*N*~190/*h* =$2^{-23}$/~185*s*/**

| $\mu \backslash \tau$ | 0 | $\tau_0/20$ | $\tau_0/10$ | $\tau_0/5$ | $\tau_0/2$ | $3\tau_0/4$ | $\tau_0$ |
|---|---|---|---|---|---|---|---|
| -1.000E+00 | 1.063698E+00 | 1.006239E+00 | 9.632064E-01 | 8.582423E-01 | 5.245333E-01 | 2.460023E-01 | 0.000000E-00 |
| -9.000E-01 | 9.530901E-01 | 9.956623E-01 | 9.697242E-01 | 8.693898E-01 | 5.359888E-01 | 2.574048E-01 | 0.000000E-00 |
| -8.000E-01 | 9.540765E-01 | 9.982827E-01 | 9.777659E-01 | 8.805282E-01 | 5.474443E-01 | 2.688357E-01 | 0.000000E-00 |
| -7.000E-01 | 8.825418E-01 | 9.885061E-01 | 9.835186E-01 | 8.915683E-01 | 5.588997E-01 | 2.802815E-01 | 0.000000E-00 |
| -6.000E-01 | 8.247123E-01 | 9.790987E-01 | 9.889036E-01 | 9.025562E-01 | 5.703552E-01 | 2.917343E-01 | 0.000000E-00 |
| -5.000E-01 | 7.726057E-01 | 9.697724E-01 | 9.939905E-01 | 9.134975E-01 | 5.818106E-01 | 3.031902E-01 | 0.000000E-00 |
| -4.000E-01 | 7.114385E-01 | 9.580045E-01 | 9.983295E-01 | 9.243758E-01 | 5.932660E-01 | 3.146475E-01 | 0.000000E-00 |
| -3.000E-01 | 6.403106E-01 | 9.434253E-01 | 1.001776E+00 | 9.351789E-01 | 6.047214E-01 | 3.261052E-01 | 0.000000E-00 |
| -2.000E-01 | 5.584817E-01 | 9.258343E-01 | 1.004214E+00 | 9.458945E-01 | 6.161767E-01 | 3.375630E-01 | 0.000000E-00 |
| -1.000E-01 | 4.587340E-01 | 9.045925E-01 | 1.005477E+00 | 9.565094E-01 | 6.276320E-01 | 3.490206E-01 | 0.000000E-00 |
| 0.000E+00 | 2.515825E-01 | 8.795200E-01 | 1.005475E+00 | 9.670143E-01 | 6.390873E-01 | 3.604781E-01 | 0.000000E-00 |
| 0.000E+00 | 0.000000E-00 | 8.795200E-01 | 1.005475E+00 | 9.670143E-01 | 6.390873E-01 | 3.604781E-01 | 3.926386E-02 |
| 1.000E-01 | 0.000000E-00 | 8.506962E-01 | 1.004213E+00 | 9.774071E-01 | 6.505425E-01 | 3.719354E-01 | 7.203907E-02 |
| 2.000E-01 | 0.000000E-00 | 8.187184E-01 | 1.001881E+00 | 9.877015E-01 | 6.619977E-01 | 3.833925E-01 | 8.894897E-02 |
| 3.000E-01 | 0.000000E-00 | 7.852151E-01 | 9.990155E-01 | 9.979409E-01 | 6.734528E-01 | 3.948494E-01 | 1.041404E-01 |
| 4.000E-01 | 0.000000E-00 | 7.545986E-01 | 9.967883E-01 | 1.008227E+00 | 6.849080E-01 | 4.063061E-01 | 1.183839E-01 |
| 5.000E-01 | 0.000000E-00 | 7.351621E-01 | 9.976028E-01 | 1.018771E+00 | 6.963633E-01 | 4.177627E-01 | 1.319904E-01 |
| 6.000E-01 | 0.000000E-00 | 7.376549E-01 | 1.005957E+00 | 1.030002E+00 | 7.078188E-01 | 4.292192E-01 | 1.451393E-01 |
| 7.000E-01 | 0.000000E-00 | 7.779241E-01 | 1.030062E+00 | 1.042782E+00 | 7.192750E-01 | 4.406756E-01 | 1.579577E-01 |
| 8.000E-01 | 0.000000E-00 | 8.870455E-01 | 1.085945E+00 | 1.058911E+00 | 7.307327E-01 | 4.521319E-01 | 1.705433E-01 |
| 9.000E-01 | 0.000000E-00 | 1.153939E+00 | 1.214127E+00 | 1.082650E+00 | 7.421936E-01 | 4.635881E-01 | 1.829543E-01 |
| 1.000E+00 | 0.000000E-00 | 8.074596E+01 | 1.178612E+01 | 1.260596E+00 | 7.536635E-01 | 4.750442E-01 | 1.952312E-01 |

gives intensities uniformly in the direction interval [-1,1] to several units in the last place indicated by emboldened digits. For doubling $l$ varied form $21 < l < 25$. A six-place benchmark is given in Table IVb confirmed to all digits by Table IVa. The computational time for these tables is 265*s* and 185*s* respectively.

## CONCLUSION

A new radiative transfer numerical solution enters into the elite category of an extreme benchmark. The method, as some are, is counterintuitive in that it takes a linear problem into a non-linear one. The centerpiece of the Matrix Riccati Equation Method (MREM) is the set of four non-linear matrix Riccati ODEs that result from the fundamental radiative transfer equation as a first order ODE. Any two of these equations provides the interaction coefficients of reflectance ***R*** and ***T*** transmittance. These coefficients are numerically found from a TS representation for a thin slab to initiate doubling. Forming the interaction principle with the interaction coefficients and doubling then enables the response for slabs of any thickness. With the slab response, the surface exiting and interior intensities become known. We have established a numerical method based on TS evaluation, doubling and adding wrapped in the Wynn-epsilon convergence acceleration to give a highly precise radiative transfer solution. Tables for the HAZE L and Cloud C1 benchmarks to nearly seven-place precision are included. All exiting intensities are precise to seven digits in comparison to RM/DOM.

## APPENDIX A

### Previous Benchmark Solutions to the 1D Radiative Transfer Equation

### A.1 The response matrix discrete ordinates method (RM/DOM) [10]

The response matrix solution is similar to the Analytical Discrete Ordinates (ADO) method [14], with one major difference. While ADO generates a solution based on the discrete form of the singular eigenfunctions leading to exponential eigenfunctions, RM/DOM assumes the even/odd parity form of the transport equation leading to modal solutions of hyperbolic type in the space of solutions of second order ODEs. Matrix diagonalization then yields two independent solutions. One advantage, apart from compactness, is that there always exists two independent solutions to the parity equations even in the case of no absorption. Thus, special attention is avoided for $\omega = 1$. In addition, the solution features the development of a response matrix (as the name suggests) over a homogeneous slab of any thickness dependent on material properties only. Responses, seamlessly added, represent contiguous heterogeneous media using the interaction principle as described above. Additional features include convergence acceleration in quadrature order through Wynn-epsilon acceleration; faux quadrature to assess precision with quadrature grid refinement also enabling a beam source to be specified in any direction; and a consistent solution at $\mu = 0$. We have achieved seven-place precision for the Cloud C1 300 term scattering kernel with quadrature order $2N$ of ~300.

### A.2 Method of doubling (MoD) [16]

This solution closely follows that developed by Van De Hulst [16] but applied to the fully discretized 1D radiative transfer equation in both space and direction. The slab is discretized and a response for a thin slice of the discrete slab is continuously doubled until the entire discrete slab is covered. The full slab is then covered by adding via the interaction principle. Richardson and Wynn-epsilon convergence acceleration applies to the initial thin slice size and to the quadrature order respectively. In addition, Richardson extrapolation is applied to the original slab discretization. The features found in RM/DOM are also included. Heterogeneous media are naturally accommodated with this approach. The Cloud C1 case is as precise as that of RM/DOM. The advantage of doubling is that no matrix diagonalization requiring eigenvalues is necessary and essentially a host of spatial discretization schemes to evaluate the matrix exponential as a solution to a first order ODE are possible. These include single and multisteps methods. The doubling solution demonstrates that the most basic of numerical schemes provide highly precise numerical solutions to the radiative transfer equation thus demonstrating simplicity.

### A.3 Double PN method (DPN) [17]

The DPN method, though theoretically equivalent to the discrete ordinates method, is not necessarily numerically equivalent. The usual half range Legendre polynomial expansion is assumed yielding moments equations closed by setting the last moment to zero. These equations form a set of ODEs solved in the even/odd parity form. Again, as in the RM/DOM, matrix diagonalization enables an explicit solution yielding a response matrix. With all the features included in the previous solutions, DPN also gives nearly seven places for the Cloud C1 case. What is particularly interesting about RM/DOM and DPN is that they both use the same solutions as found in solving the diffusion equation, which gives another connection between diffusion and transport theory. This is not surprising since they are all solving second order ODEs.

## APPENDIX B

## Indirect Taylor Series Representation

### B.1 Reflectance

One can linearize Eq(12c) with the following transformation:

$$\boldsymbol{R}(\tau) \equiv \boldsymbol{Q}(\tau)\boldsymbol{P}^{-1}(\tau) \tag{B1}$$

as shown. Since from Eq(B1) in Eq(12c), by differentiation and grouping, one finds

$$\frac{d\boldsymbol{Q}}{d\tau}\boldsymbol{P}^{-1} + \boldsymbol{Q}\frac{d\boldsymbol{P}^{-1}}{d\tau} = [\boldsymbol{\beta P} - \boldsymbol{\alpha Q}]\boldsymbol{P}^{-1} - \boldsymbol{QP}^{-1}\boldsymbol{\alpha} + \boldsymbol{QP}^{-1}\boldsymbol{\beta QP}^{-1}. \tag{B2}$$

If one defines

$$\frac{d\boldsymbol{Q}}{d\tau} \equiv \boldsymbol{\beta P} - \boldsymbol{\alpha Q}, \tag{B3}$$

and pre-multiplies by $\boldsymbol{P}$, then

$$\boldsymbol{P}\frac{d\boldsymbol{P}^{-1}}{d\tau} = -\boldsymbol{\alpha} + \boldsymbol{\beta QP}^{-1}. \tag{B4}$$

Noting

$$\frac{d\boldsymbol{P}\boldsymbol{P}^{-1}}{d\tau} = \frac{d\boldsymbol{I}}{d\tau} = 0 = \boldsymbol{P}\frac{d\boldsymbol{P}^{-1}}{d\tau} + \frac{d\boldsymbol{P}}{d\tau}\boldsymbol{P}^{-1},$$

Eq(B4) becomes

$$-\frac{d\boldsymbol{P}}{d\tau}\boldsymbol{P}^{-1} = -\boldsymbol{\alpha} + \boldsymbol{\beta}\boldsymbol{Q}\boldsymbol{P}^{-1}$$

to give

$$\frac{d\boldsymbol{P}}{d\tau} = \boldsymbol{\alpha}\boldsymbol{P} - \boldsymbol{\beta}\boldsymbol{Q}. \qquad \text{(B5)}$$

If

$$\boldsymbol{Y} \equiv \begin{bmatrix} \boldsymbol{P} \\ \boldsymbol{Q} \end{bmatrix}, \qquad \text{(B6)}$$

Eqs (B3) and (B5) combine as

$$\frac{d\boldsymbol{Y}}{d\tau} \equiv \boldsymbol{A}\boldsymbol{Y} \qquad \text{(B7a)}$$

with initial condition from

$$\boldsymbol{R}(0) \equiv \boldsymbol{Q}(0)\boldsymbol{P}^{-1}(0) = 0$$

satisfied by

$$\begin{aligned} &\boldsymbol{Q}(0) \equiv \boldsymbol{0} \\ &\boldsymbol{P}^{-1}(0) = \boldsymbol{P}(0) \equiv \boldsymbol{I}_N \end{aligned}$$

to imply

$$\boldsymbol{Y}(0) \equiv \begin{bmatrix} \boldsymbol{I}_N \\ \boldsymbol{0} \end{bmatrix}. \qquad \text{(B7b)}$$

To find the components of $\boldsymbol{Y}(\tau)$, we impose the TS

$$\boldsymbol{Y}(\tau)=\sum_{n=0}^{\infty}\boldsymbol{Y}_n\tau^n \tag{B8a}$$

to give

$$\frac{1}{n!}\boldsymbol{Y}^{(n)}(\tau)\Big|_{\tau=0}\equiv\boldsymbol{Y}_n=\frac{1}{n}\boldsymbol{A}\boldsymbol{Y}_{n-1}. \tag{B8b}$$

and therefore from Eq(B7a)

$$\begin{aligned}\frac{1}{n!}\boldsymbol{P}^{(n)}(\tau)\Big|_{\tau=0}&\equiv\boldsymbol{P}_n=\frac{1}{n}\left[\boldsymbol{\alpha}\boldsymbol{P}_{n-1}-\boldsymbol{\beta}\boldsymbol{Q}_{n-1}\right]\\ \frac{1}{n!}\boldsymbol{Q}^{(n)}(\tau)\Big|_{\tau=0}&\equiv\boldsymbol{Q}_n=\frac{1}{n}\left[\boldsymbol{\beta}\boldsymbol{P}_{n-1}-\boldsymbol{\alpha}\boldsymbol{Q}_{n-1}\right].\end{aligned} \tag{B9}$$

For the TS coefficients for $\boldsymbol{R}$ of Eq(B1), we write

$$\boldsymbol{P}(\tau)\boldsymbol{R}(\tau)\equiv\boldsymbol{Q}(\tau), \tag{B10}$$

and substituting TS

$$\boldsymbol{P}_0\boldsymbol{R}_n+\sum_{l=1}^{n}\boldsymbol{P}_{n-l}\boldsymbol{R}_l=\boldsymbol{Q}_n; \tag{B11}$$

and therefore the indirect TS coefficients for the reflection coefficient

$$\boldsymbol{R}_n=\boldsymbol{Q}_n-\sum_{l=1}^{n}\boldsymbol{P}_{n-l}\boldsymbol{R}_l\,. \tag{B12}$$

**B.2 Transmittance**

For transmission, we begin with Eq()

$$\frac{d\boldsymbol{T}}{d\tau}=\boldsymbol{R}\boldsymbol{\beta}\boldsymbol{T}-\boldsymbol{\alpha}\boldsymbol{T} \tag{B13}$$

into which we introduce

$$\boldsymbol{T} = \boldsymbol{W}\boldsymbol{P}^{-1}, \tag{B14}$$

to find

$$\begin{aligned}\frac{d\boldsymbol{T}}{d\tau} &= \frac{d\boldsymbol{W}}{d\tau}\boldsymbol{P}^{-1} + \boldsymbol{W}\frac{d\boldsymbol{P}^{-1}}{d\tau} \\ &= \boldsymbol{Q}\boldsymbol{P}^{-1}\boldsymbol{\beta}\boldsymbol{W}\boldsymbol{P}^{-1} - \boldsymbol{\alpha}\boldsymbol{W}\boldsymbol{P}^{-1}.\end{aligned} \tag{B15}$$

Since

$$\begin{aligned}&\frac{d\boldsymbol{W}}{d\tau}\boldsymbol{P}^{-1} + \boldsymbol{W}\frac{d\boldsymbol{P}^{-1}}{d\tau} \\ &= \boldsymbol{Q}\boldsymbol{P}^{-1}\boldsymbol{\beta}\boldsymbol{W}\boldsymbol{P}^{-1} - \boldsymbol{\alpha}\boldsymbol{W}\boldsymbol{P}^{-1}\end{aligned} \tag{B16a}$$

and from Eq(B4) above

$$\frac{d\boldsymbol{P}^{-1}}{d\tau} = -\boldsymbol{P}^{-1}\boldsymbol{\alpha} + \boldsymbol{P}^{-1}\boldsymbol{\beta}\boldsymbol{Q}\boldsymbol{P}^{-1}, \tag{B16b}$$

there results

$$\begin{aligned}&\frac{d\boldsymbol{W}}{d\tau}\boldsymbol{P}^{-1} - \boldsymbol{W}\boldsymbol{P}^{-1}\boldsymbol{\alpha} + \boldsymbol{W}\boldsymbol{P}^{-1}\boldsymbol{\beta}\boldsymbol{Q}\boldsymbol{P}^{-1} = \\ &= \boldsymbol{Q}\boldsymbol{P}^{-1}\boldsymbol{\beta}\boldsymbol{W}\boldsymbol{P}^{-1} - \boldsymbol{\alpha}\boldsymbol{W}\boldsymbol{P}^{-1}.\end{aligned} \tag{B16c}$$

From the second identity of Eq(12f)

$$\boldsymbol{Q}\boldsymbol{P}^{-1}\boldsymbol{\beta}\boldsymbol{W}\boldsymbol{P}^{-1} - \boldsymbol{\alpha}\boldsymbol{W}\boldsymbol{P}^{-1} = -\boldsymbol{W}\boldsymbol{P}^{-1}\boldsymbol{\alpha} + \boldsymbol{W}\boldsymbol{P}^{-1}\boldsymbol{\beta}\boldsymbol{Q}\boldsymbol{P}^{-1}$$

resulting in

$$\frac{d\boldsymbol{W}}{d\tau} = 0. \tag{B17a}$$

Thus,

$$\boldsymbol{W}(\tau) = \boldsymbol{C} \tag{B17b}$$

and since

$$\boldsymbol{T}(0) = \boldsymbol{I}_N = \boldsymbol{W}(0)\boldsymbol{P}^{-1}(0) = \boldsymbol{W}(0)\boldsymbol{I}_N$$

and

$$\boldsymbol{W}(\tau) = \boldsymbol{W}(0) = \boldsymbol{I}_N . \tag{B17c}$$

The TS emerges when Eq(B14) is re-expressed as

$$\boldsymbol{P}(\tau)\boldsymbol{T}(\tau) = \boldsymbol{I}_N \tag{B18}$$

to give the TS coefficients

$$\sum_{l=0}^{n} \boldsymbol{P}_{n-l}\boldsymbol{T}_l = \boldsymbol{I}_N \delta_{n0}$$

and therefore

$$\boldsymbol{T}_n = \boldsymbol{I}_N \delta_{n0} - \sum_{l=0}^{n-1} \boldsymbol{P}_{n-l}\boldsymbol{T}_l . \tag{B19}$$

and

$$\boldsymbol{T} = \boldsymbol{P}^{-1}. \tag{B20}$$

One finds the identical result if Eq(12b) were used as the initial equation rather than Eq(12d).

Seems, the last equation is new and implies

$$\boldsymbol{R}(\tau) = \boldsymbol{Q}(\tau)\boldsymbol{T}(\tau). \tag{B21}$$

Additional implications of this expression needs further study in particular for the eigenvalue based radiative transfer formulation [10].

## APPENDIX C

### Legendre Coefficients for Scattering Phase Functions

The following are the scattering coefficients for HAZE L and Cloud C1 phase functions represented to orders $L$ =82 and 300 respectively as the Legendre series

$$f(\mu,\mu') \equiv \frac{1}{2}\sum_{l=0}^{L} \beta_l P(\mu) P(\mu'). \tag{C.1}$$

## C.1 HAZE L

| $l$ | $\beta_l$ | $\beta_{l+15}$ | $\beta_{l+30}$ | $\beta_{l+45}$ | $\beta_{l+60}$ | $\beta_{l+75}$ |
|---|---|---|---|---|---|---|
| 0 | 1.00000 | 0.42563 | 0.02451 | 0.00160 | 0.00015 | 0.00001 |
| 1 | 2.41260 | 0.34688 | 0.01874 | 0.00150 | 0.00012 | 0.00001 |
| 2 | 3.23047 | 0.28351 | 0.01711 | 0.00115 | 0.00011 | 0.00001 |
| 3 | 3.37296 | 0.23317 | 0.01298 | 0.00107 | 0.00009 | 0.00001 |
| 4 | 3.23150 | 0.18963 | 0.01198 | 0.00082 | 0.00008 | 0.00001 |
| 5 | 2.89350 | 0.15788 | 0.00904 | 0.00077 | 0.00006 | 0.00001 |
| 6 | 2.49594 | 0.12739 | 0.00841 | 0.00059 | 0.00006 | 0.00001 |
| 7 | 2.11361 | 0.10762 | 0.00634 | 0.00055 | 0.00005 | 0.00001 |
| 8 | 1.74812 | 0.08597 | 0.00592 | 0.00043 | 0.00004 | |
| 9 | 1.44692 | 0.07381 | 0.00446 | 0.00040 | 0.00004 | |
| 10 | 1.17714 | 0.05828 | 0.00418 | 0.00031 | 0.00003 | |
| 11 | 0.96643 | 0.05089 | 0.00316 | 0.00029 | 0.00003 | |
| 12 | 0.78237 | 0.03971 | 0.00296 | 0.00023 | 0.00002 | |
| 13 | 0.64114 | 0.03524 | 0.00225 | 0.00021 | 0.00002 | |
| 14 | 0.51966 | 0.02720 | 0.00210 | 0.00017 | 0.00002 | |

## C.2 Cloud C1

| $l$ | $\beta_l$ | $\beta_{l+30}$ | $\beta_{l+60}$ | $\beta_{l+90}$ | $\beta_{l+120}$ | $\beta_{l+150}$ | $\beta_{l+180}$ | $\beta_{l+210}$ | $\beta_{l+240}$ | $\beta_{l+270}$ |
|---|---|---|---|---|---|---|---|---|---|---|
| 0 | 1.000 | 18.885 | 18.551 | 10.618 | 4.285 | 1.344 | 0.349 | 0.079 | 0.016 | 0.003 |
| 1 | 2.544 | 19.103 | 18.348 | 10.35 | 4.130 | 1.284 | 0.331 | 0.074 | 0.015 | 0.003 |
| 2 | 3.883 | 19.345 | 18.119 | 10.09 | 3.994 | 1.235 | 0.317 | 0.071 | 0.014 | 0.003 |
| 3 | 4.568 | 19.537 | 17.901 | 9.827 | 3.847 | 1.179 | 0.301 | 0.067 | 0.013 | 0.002 |
| 4 | 5.235 | 19.721 | 17.659 | 9.574 | 3.719 | 1.134 | 0.288 | 0.064 | 0.013 | 0.002 |
| 5 | 5.887 | 19.884 | 17.428 | 9.318 | 3.580 | 1.082 | 0.273 | 0.060 | 0.012 | 0.002 |
| 6 | 6.457 | 20.024 | 17.174 | 9.072 | 3.459 | 1.040 | 0.262 | 0.057 | 0.011 | 0.002 |
| 7 | 7.177 | 20.145 | 16.931 | 8.822 | 3.327 | 0.992 | 0.248 | 0.054 | 0.011 | 0.002 |
| 8 | 7.859 | 20.251 | 16.668 | 8.584 | 3.214 | 0.954 | 0.238 | 0.052 | 0.010 | 0.002 |
| 9 | 8.494 | 20.330 | 16.415 | 8.340 | 3.090 | 0.909 | 0.225 | 0.049 | 0.009 | 0.002 |
| 10 | 9.286 | 20.401 | 16.144 | 8.110 | 2.983 | 0.873 | 0.215 | 0.047 | 0.009 | 0.002 |
| 11 | 9.856 | 20.444 | 15.883 | 7.874 | 2.866 | 0.832 | 0.204 | 0.044 | 0.008 | 0.002 |
| 12 | 10.615 | 20.477 | 15.606 | 7.652 | 2.766 | 0.799 | 0.195 | 0.042 | 0.008 | 0.001 |
| 13 | 11.229 | 20.489 | 15.338 | 7.424 | 2.656 | 0.762 | 0.185 | 0.039 | 0.008 | 0.001 |
| 14 | 11.851 | 20.483 | 15.058 | 7.211 | 2.562 | 0.731 | 0.177 | 0.038 | 0.007 | 0.001 |
| 15 | 12.503 | 20.467 | 14.784 | 6.990 | 2.459 | 0.696 | 0.167 | 0.035 | 0.007 | 0.001 |
| 16 | 13.058 | 20.427 | 14.501 | 6.785 | 2.372 | 0.668 | 0.160 | 0.034 | 0.006 | 0.001 |
| 17 | 13.626 | 20.382 | 14.225 | 6.573 | 2.274 | 0.636 | 0.151 | 0.032 | 0.006 | 0.001 |
| 18 | 14.209 | 20.310 | 13.941 | 6.377 | 2.193 | 0.610 | 0.145 | 0.030 | 0.006 | 0.001 |
| 19 | 14.660 | 20.236 | 13.662 | 6.173 | 2.102 | 0.581 | 0.137 | 0.029 | 0.005 | 0.001 |
| 20 | 15.231 | 20.136 | 13.378 | 5.986 | 2.025 | 0.557 | 0.131 | 0.027 | 0.005 | 0.001 |
| 21 | 15.641 | 20.036 | 13.098 | 5.790 | 1.940 | 0.530 | 0.124 | 0.026 | 0.005 | 0.001 |
| 22 | 16.126 | 19.909 | 12.816 | 5.612 | 1.869 | 0.508 | 0.118 | 0.024 | 0.005 | 0.001 |
| 23 | 16.539 | 19.785 | 12.536 | 5.424 | 1.790 | 0.483 | 0.112 | 0.023 | 0.004 | 0.001 |
| 24 | 16.934 | 19.632 | 12.257 | 5.255 | 1.723 | 0.463 | 0.107 | 0.022 | 0.004 | 0.001 |
| 25 | 17.325 | 19.486 | 11.978 | 5.075 | 1.649 | 0.440 | 0.101 | 0.021 | 0.004 | 0.001 |
| 26 | 17.673 | 19.311 | 11.703 | 4.915 | 1.588 | 0.422 | 0.097 | 0.020 | 0.004 | 0.001 |
| 27 | 17.999 | 19.145 | 11.427 | 4.744 | 1.518 | 0.401 | 0.091 | 0.018 | 0.003 | 0.001 |
| 28 | 18.329 | 18.949 | 11.156 | 4.592 | 1.461 | 0.384 | 0.087 | 0.018 | 0.003 | 0.001 |
| 29 | 18.588 | 18.764 | 10.884 | 4.429 | 1.397 | 0.364 | 0.082 | 0.017 | 0.003 | 0.001 |